\documentclass[acmsmall, screen=true, review=false]{acmart}

\acmJournal{TRETS}
\acmVolume{1}
\acmNumber{1}
\acmArticle{1}
\acmYear{2019}
\acmMonth{5}
\acmPrice{15.00}
\copyrightyear{2019}
\setcopyright{acmcopyright}
\acmDOI{10.1145/3337929}

\usepackage{mathtools}
\DeclarePairedDelimiter{\ceil}{\lceil}{\rceil}
\usepackage{algorithm}
\usepackage[noend]{algpseudocode}
\usepackage{pgfplots}
\pgfplotsset{compat=newest}
\usepgfplotslibrary{colorbrewer}
\pgfplotsset{cycle list/Dark2-8}
\usepackage{pgfplotstable}
\usepackage{booktabs}
\usepackage{multirow}
\usepackage{todonotes}
\usepackage{tikz}
\usetikzlibrary{circuits.logic.US}

\definecolor{LHS}{RGB}{0, 80, 158} %121, 162, 206}
\definecolor{RHS}{RGB}{92, 190, 201} %85, 41, 136}

\graphicspath{{figures/}}
\DeclareGraphicsExtensions{.pdf}

\newcommand{\OurScheme}{\textsc{BISMO}}

\begin{document}

\title{Optimizing Bit-Serial Matrix Multiplication for Reconfigurable Computing}

\author{Yaman Umuroglu}
\affiliation{%
  \institution{Xilinx Research Labs}
  \city{Dublin}
  \country{Ireland}
}
\email{yamanu@xilinx.com}

\author{Davide Conficconi}
\affiliation{%
  \institution{Xilinx Research Labs}
  \city{Dublin}
  \country{Ireland}
}
\affiliation{%
  \institution{Politecnico di Milano}
  \city{Milano}
  \country{Italy}
}
\email{davidec@xilinx.com}

\author{Lahiru Rasnayake}
\affiliation{%
  \institution{Norwegian University of Science and Technology}
  \city{Trondheim}
  \country{Norway}
}
\email{lahiru.rasnayake@ntnu.no}

\author{Thomas B. Preusser}
\affiliation{%
  \institution{Accemic Technologies GmbH}
  \city{Dresden}
  \country{Germany}
}
\email{thomas.preusser@utexas.edu}

\author{Magnus Sj\"alander}
\orcid{0003-4232-6976}
\affiliation{%
  \institution{Uppsala University}
  \city{Uppsala}
  \country{Sweden}
}
\affiliation{%
  \institution{Norwegian University of Science and Technology}
  \city{Trondheim}
  \country{Norway}
}
\email{magnus.sjalander@ntnu.no}

\begin{abstract}
  % High-precision computation is costly in terms of compute resources, runtime,
  % and energy.
  Matrix-matrix multiplication is a key computational kernel for numerous
  applications in science and engineering, with ample parallelism and data
  locality that lends itself well to high-performance implementations. %
  Many matrix multiplication-dependent applications can use reduced-precision
  integer or fixed-point representations to increase their performance and
  energy efficiency while still offering adequate quality of results. %
  % An application may require different precisions at different stages of its
  % execution or the precision might depend on the input data,
  % \textcolor{blue}{Alternatively: might be input data dependent} rendering
  % constant-precision solutions ineffective. %
  However, precision requirements may vary between different application phases
  or depend on input data, rendering constant-precision solutions
  ineffective.  %
  \OurScheme{}, a vectorized bit-serial matrix multiplication overlay for
  reconfigurable computing, previously utilized the excellent binary-operation
  performance
  % binary matrix multiplication capabilities
  of FPGAs to offer a matrix multiplication performance that scales with
  required precision and parallelism. %
  We show how \OurScheme{} can be scaled up on Xilinx FPGAs using an
  arithmetic architecture that better utilizes 6-input LUTs.
  % , and further enhanced by a data layout converter for faster end-to-end
  % operation.
  The improved \OurScheme{} achieves a peak performance of 15.4 binary TOPS on
  the Ultra96 board with a Xilinx UltraScale+~MPSoC.
  % We further demonstrate the scalability of the core datapath by successfully
  % synthesizing a 783~binary TOPS array on a Xilinx VU9P.
\end{abstract}

%  LocalWords:  textcolor

%
% The code below should be generated by the tool at
% http://dl.acm.org/ccs.cfm
% Please copy and paste the code instead of the example below.
%



\maketitle

\renewcommand{\shortauthors}{Y. Umuroglu, D. Conficconi, L. Rasnayake, T. B. Preusser, and M. Sj\"alander}

\section{Introduction}

% First the helicopter view
Using constant precision for all operations is the predominant practice when
designing digital systems, since logical and arithmetic operations, registers,
memories, and interconnects can be designed to accommodate one specific
precision. %
% Disadvantage of fixed precision
Their main disadvantage is the associated overhead in storing, communicating,
and performing operations with full precision when an application only requires
a fraction of the supported precision. %
% Many applications do not require the full precision
Numerous applications, in the engineering, scientific, and multimedia
domain, can use reduced precision and still produce adequate results. %
% Approximate computing and Neural Networks
This property has been leveraged in approximate
computing~\cite{mittal+:CSUR2016approx_survey} and quantized neural
networks (QNNs)~\cite{hubara2016quantized, park2017weighted},
to improve performance and energy efficiency and to reduce
area by tailoring computations to the required precision. %
The required precision may vary between different phases of the application.
As an example, Park et al.~\cite{park2017weighted} achieve the best
performance-accuracy tradeoff for QNNs by using fewer bits for the intermediate
layers, and Wang et al.~\cite{wang2018haq} use a reinforcement learning approach
to discover efficient QNNs with different per-layer quantization.

\looseness -1
% Matrix multiplication and their importance
Matrix-matrix multiplication is a commonly used computational kernel and
represents one of the seven Berkeley dwarfs, which are important computational
constructs for engineering and scientific
computing~\cite{asanovic+:dwarfs-2006}. %
% Regularity and amount of computations
The amount of computation required for matrix multiplications makes it highly
beneficial to adapt the operational precision to an application's
requirements. %
% Good fit for FPGAs low-precision
FPGAs are a good fit for low-precision operations and for instantiating
efficient matrix multiplication accelerators with a specific precision. %
% Variable precision
However, fixed-precision accelerators are not suitable for applications with
variable precision as they either require multiple instances of the same
accelerator, each with a different precision,
% Dynamic reconfiguration
or require dynamic reconfiguration with associated overhead and system
complexity. %

% Bit-serial
A promising alternative to fixed-precision accelerators is to use bit-serial
computations~\cite{umuroglu_jahre:CASES2017} where the integer matrix
multiplication is expressed as a weighted sum of binary matrix multiplications
(\autoref{sec:background}). %
The bit-serial alternative provides the possibility to use one efficient binary
matrix multiplication accelerator to compute matrix multiplications of any
precision. %

% BISMO
Towards this end, a bit-serial matrix multiplication overlay called BISMO was
presented by Umuroglu et\,al.~\cite{umuroglu+:FPL2018}.
BISMO consists of a software-programmable weighted binary
matrix multiplication engine and associated hardware for fetching data and
storing back the result (\autoref{sec:hw_architecture}). %
% Configurability and cost model
The hardware architecture is design-time configurable and comes with a cost
model for estimating the resource usage for a given set of parameters
(\autoref{sec:cost_model}).
% Software scalability
BISMO's software programmability enables it to operate on any matrix size and at
any fixed-point or integer precision (\autoref{sec:sw_stack}). %

This article proposes several improvements to the original
BISMO~\cite{umuroglu+:FPL2018}.
% New compressor
We present a new and highly LUT-efficient compressor architecture for
performing the core \textsc{And}-popcount operation for bit-serial
(\autoref{sec:compressor}). %
% Improved DPU
The DPU architecture has been further improved by eliminating the need for a
barrel shifter. %
This is achieved by organizing the bit-serial matrix multiplications into
wavefronts starting with the highest weighted matrix multiplication being
performed followed by consecutively less weighted matrix multiplications
(\autoref{sec:new_DPU}). %
The new wavefront schedule requires only a fixed left shift of 1-bit instead of
a variable shift-amount. %
The new DPU architecture is compared against our previously proposed DPU
architecture ~\cite{umuroglu+:FPL2018} (\autoref{sec:synthesis_DPU}). %
% P2S accelerator
To address data layout conversion challenges for bit-serial, we introduce a
new parallel-to-serial (P2S) accelerator that takes a conventional
bit-parallel matrix and produces the equivalent bit-serial matrices
\autoref{sec:p2s_accelerator}, and evaluate its resource cost and performance
(\autoref{sec:p2s_performance}). %
% Updated cost model
We also present an updated \OurScheme{} cost model (\autoref{sec:cost_model})
that has been validated on an Ultra96 FPGA board
(\autoref{sec:synthesis_cost_model}). % ~\cite{ultra96}

% Performance and energy results
The most recent \OurScheme{} prototype achieves a top performance of 15.4 binary
TOPS at 2.1 TOPS/W power efficiency when implemented on an Ultra96 board
(\autoref{sec:performance}). %
A scalability evaluation shows that the \OurScheme{} dot product array (DPA) is
capable of achieving a peak performance of at least 783 binary TOPS on a Xilinx
Virtex UltraScale+ VU9P (\autoref{sec:scaling_DPA}). %
% while the complete BISMO accelerator achieves a peak performance of at least
% \textcolor{red}{XXX}

% \vspace{10mm}

% This article is an extended version of our previous conference paper
% \cite{umuroglu+:FPL2018}.
% In this work, we improve upon BISMO with the following contributions:
% \begin{itemize}
%   \item Improved compressor architecture optimized for Xilinx FPGAs and quantification.
%     %% This "quantification" does not read right.
%     %% Should it be something like "optimized and quantified/characterized for Xilinx devices"?
%   \item Improved DPU datapath that avoids barrel shifting and simplifies control.
%   \item New data layout and parallel-to-serial accelerator.
%   \item Synthesis results for large \OurScheme{} arrays with a maximum performance of 783 binary TOPS.
%   \item Prototypes and a new cost model on Ultra96, achieving up to 15.4 binary TOPS at 2.1 TOPS/W power efficiency.
% \end{itemize}

% \TODO{summarize key metrics}
% \OurScheme{} is open-sourced at https://git.io/fWb0m~\cite{BISMO-git:EECS2018}.

%  LocalWords:  textcolor datapath textbf autoref sec:sw_stack textit textsc
%  LocalWords:  asanovic umuroglu_jahre:CASES2017 CSUR2016approx_survey Virtex
%  LocalWords:  hubara2016quantized sec:relatedwork Umuroglu umuroglu popcount
%  LocalWords:  Zynq vspace wang2018haq

\section{Background: Bit-Serial}
\label{sec:background}

% Fixed precision arithmetic
% Creating fast parallel arithmetic circuits are challenging and costly in terms
% of resources, e.g., fast adders require complex carry-propagation networks and
% parallel multipliers require large partial-product reduction trees. %
% Designing arithmetic circuits that are fast and energy efficient while able to
% operate on variable precision inputs becomes extremely difficult. %

% Fixed precision
Fixed-precision operations have to be designed to accommodate the
largest supported precision, which causes overheads in cases where the
required precision of an application varies throughout its execution
or when the precision depends on its input data. %
% Bit serial is frugal but long-latency
In contrast, bit-serial operations are inherently frugal since they
only compute as many bits as specified by the precision of the
operands. %
However, their serial nature causes high latencies and potentially poor
performance. %
In this section, we will describe how bit-serial matrix multiplication works
on an algorithmic level, and briefly cover the data layout implications for
bit-serial matrix multiplication for implementation purposes.

\subsection{Bit-Serial Matrix Multiplication}

Matrix multiplication is a suitable kernel for taking advantage of the
frugality of bit-serial operations while overcoming the high-latency
by performing many bit-serial operations in parallel. %
Umuroglu and Jahre showed that by expressing a matrix multiplication
as a weighted sum of binary matrix multiplications
(Algorithm~\ref{alg:bit-serial_matrix_multiplication}) it is possible
to efficiently compute matrix multiplications of variable precision
using the logical \textsc{And} and population count (popcount)
instructions available in most modern
processors~\cite{umuroglu_jahre:CASES2017}. %
In addition, the algorithm works for both integer as well as fixed point
number representations, where the new fixed point location is given by the
product of the input matrices' scaling factors. %

\begin{algorithm}[t]
  \small
    \centering
    \begin{algorithmic}[1]
      \algrenewcommand\algorithmicindent{1.0em}
      \State \textbf{Input:} $M \times K$ $l$-bit matrix $L$, $K \times N$ $r$-bit matrix $R$
      \State \textbf{Output:} $P = L \cdot R$
      \For{$i \gets 0 \dots l-1$}
        \For{$j \gets 0 \dots r-1$}
          \State $\mathrm{sgnL} \gets (i == l-1 \mathrel{?} -1 : 1)$
          \State $\mathrm{sgnR} \gets (j == r-1 \mathrel{?} -1 : 1)$
          \State $\mathrm{weight} = \mathrm{sgnL} \cdot \mathrm{sgnR} \cdot
          2^{i+j}$
          \State \textit{\#  Binary matrix multiplication between
            $L^{[i]}$ and $R^{[j]}$ }
          \State \textit{\# $L^{[i]}_{mk}$ refers to $i^\mathrm{th}$ bit position of
            element at row $m$ and column $k$ of matrix $L$}
          \For{$m \gets 1 \dots M$}
            \For{$n \gets 1 \dots N$}
                \For{$k \gets 1 \dots K$}
                    \State $P_{mn} = P_{mn} +  \mathrm{weight} \cdot (L^{[i]}_{mk} \cdot R^{[j]}_{kn}) $
                \EndFor
            \EndFor
        \EndFor
        \EndFor
      \EndFor
    \end{algorithmic}
    \caption{\small Bit-serial matrix multiplication on signed two's complement integers.}
    \label{alg:bit-serial_matrix_multiplication}
\end{algorithm}

\autoref{fig:bit-serial_example} illustrates
Algorithm~\ref{alg:bit-serial_matrix_multiplication} for the example
where the two input-matrices ($L$ and $R$) consist of 2-bit unsigned
integer numbers. %
By expressing $L$ and $R$ as weighted sums of binary matrices, the
matrix product ($P = L\cdot R$) can be expressed as a weighted sum of
products between binary matrices. %
The matrix multiplication can thus be expressed as a large number of
binary operations that can be performed in parallel. %

\begin{figure}[th]
\begin{equation*}
\begin{aligned}
  L = &
  \begin{bmatrix}
    2 & 0 \\
    1 & 3
  \end{bmatrix}
  = 2^1 \textcolor{blue}{L^{[1]}} + 2^0 \textcolor{red}{L^{[0]}}
  = 2^1
%  \times
  {
  \color{blue}
  \begin{bmatrix}
    1 & 0 \\
    0 & 1
  \end{bmatrix}
  }
  + 2^0
%  \times
  {
  \color{red}
  \begin{bmatrix}
    0 & 0 \\
    1 & 1
  \end{bmatrix}
  }
  \\
  R = &
  \begin{bmatrix}
    0 & 1 \\
    1 & 2
  \end{bmatrix}
  = 2^1 \textcolor{blue}{R^{[1]}} + 2^0 \textcolor{red}{R^{[0]}}
  = 2^1
%  \times
  {
  \color{blue}
  \begin{bmatrix}
    0 & 0 \\
    0 & 1
  \end{bmatrix}
  }
  + 2^0
%  \times
  {
  \color{red}
  \begin{bmatrix}
    0 & 1 \\
    1 & 0
  \end{bmatrix}
  }
  \\
  P = &
  L \cdot R =
  (2^1 \textcolor{blue}{L^{[1]}} + 2^0 \textcolor{red}{L^{[0]}})
  \cdot
  (2^1 \textcolor{blue}{R^{[1]}} + 2^0 \textcolor{red}{R^{[0]}}) \\
  = &
  2^2 \textcolor{blue}{L^{[1]}} \cdot \textcolor{blue}{R^{[1]}}
  +
  2^1 \textcolor{blue}{L^{[1]}} \cdot \textcolor{red}{R^{[0]}} +
  2^1 \textcolor{red}{L^{[0]}} \cdot \textcolor{blue}{R^{[1]}}
  +
  2^0 \textcolor{red}{L^{[0]}} \cdot \textcolor{red}{R^{[0]}} \\
%   = &
%   2^2
% %  \times
%   {
%   \color{blue}
%   \begin{bmatrix}
%     1 & 0 \\
%     0 & 1
%   \end{bmatrix}
%   }
%   \cdot
%   {
%   \color{blue}
%   \begin{bmatrix}
%     0 & 0 \\
%     0 & 1
%   \end{bmatrix}
%   }
%   \\
%   + &
%   2^1
% %  \times
% %  \left(
%   {
%   \color{blue}
%   \begin{bmatrix}
%     1 & 0 \\
%     0 & 1
%   \end{bmatrix}
%   }
%   \cdot
%   {
%   \color{red}
%   \begin{bmatrix}
%     0 & 1 \\
%     1 & 0
%   \end{bmatrix}
%   }
%   + 2^1
% %  \times
%   {
%   \color{red}
%   \begin{bmatrix}
%     0 & 0 \\
%     1 & 1
%   \end{bmatrix}
%   }
%   \cdot
%   {
%   \color{blue}
%   \begin{bmatrix}
%     0 & 0 \\
%     0 & 1
%   \end{bmatrix}
%   }
% %  \right)
%   \\
%   + &
%   2^0
% %  \times
%   {
%   \color{red}
%   \begin{bmatrix}
%     0 & 0 \\
%     1 & 1
%   \end{bmatrix}
%   }
%   \cdot
%   {
%   \color{red}
%   \begin{bmatrix}
%     0 & 1 \\
%     1 & 0
%   \end{bmatrix}
%   }
\end{aligned}
\end{equation*}
\caption{Example of a bit-serial matrix multiplication on unsigned
  integers % with the two first for-loops unrolled
  (Algorithm~\ref{alg:bit-serial_matrix_multiplication}: for-loop on
  line 3 and 4 unrolled and weight on line 7 always positive).}
\label{fig:bit-serial_example}
\end{figure}

\subsection{Bit-Serial Data Layout}
\label{sec:bsdatalayout}
From an implementation point of view, it is important to match the data delivered
by the memory system of an accelerator and what the algorithm implemented by
the accelerator expects.
Typically, the memory system will deliver a number of bits grouped together in
response to a request.
If the order of bits provided by the memory is substantially different from the
order in which the accelerator expects them, the memory bandwidth will be
underutilized.
For bit-serial matrix multiplication, the data layout requirements are
substantially different than bit-parallel matrix multiplication.
A bit-parallel layout, where all bit positions of an element are consecutive,
is well-matched with bit-parallel matrix multiplication, which makes use of all
bit positions at once.
In contrast, bit-serial works on a single bit position at a time, but
the same bit position for neighboring elements can be processed together.
If the input matrices are provided in bit-parallel format, they should first be
converted into a bit-serial layout to ensure performance.

In this work, we assume the \textbf{[bits][rows][columns]} data layout for
bit-serial matrices, as was also assumed in prior work
\cite{umuroglu_jahre:CASES2017, pedersoli2017espresso, umuroglu+:FPL2018}. %
\autoref{sec:p2s_accelerator} provides an example of this data layout in context
of a parallel-to-serial accelerator for BISMO. %

%  LocalWords:  textcolor umuroglu_jahre:CASES2017 bmatrix algrenewcommand
%  LocalWords:  algorithmicindent textbf cdot alg:bs_matrix_mul sgnR
%  LocalWords:  mathrm mathrel sgnL textit rc Umuroglo textsc autoref
%  LocalWords:  alg:bit-serial_matrix_multiplication popcount mn mk
%  LocalWords:  Umuroglu sec:bsdatalayout pedersoli2017espresso

\section{The Bit-Serial Matrix Multiplication Overlay}

\OurScheme{} consists of a hardware part and a software part. %
% HW
The hardware part is composed of a scalable bit-serial matrix
multiplication datapath and associated memory and control logic. %
% SW
The software part generates instructions for the hardware for a given
matrix size and precision. %
The key features offered by this hardware-software design are the
following: %

\textbf{Precision Scalability.} %
By expressing an integer or fixed-point matrix multiplication as a
weighted sum of binary matrix multiplications
(\autoref{sec:background}), the same hardware can be utilized for a
range of different precisions. %
Lower-precision matrix multiplications are finished quickly, while
higher-precision requires more clock cycles. %

\textbf{Hardware Scalability.} %
Our overlay generator can scale the memory and compute resource
utilization to match system-level requirements. %
This is achieved by controlling the parameters described in
\autoref{sec:hw_architecture}. %
% \textbf{Dot Product Unit}
The dot product unit (DPU) is \OurScheme{}'s core processing element,
which performs a multiply and accumulate between two weighted binary
vectors. %
We present a new DPU datapath and an efficient FPGA compressor
(\autoref{sec:DPU}) that improves resource utilization and DPU
scalability. %
% \textbf{Parallel-to-Serial Data Format Transformation.} %
A parallel-to-serial (P2S) accelerator is described in
\autoref{sec:p2s_accelerator}, which takes bit-parallel matrices and
transforms them into the required bit-serial data format. %
% \textbf{Cost model.} %
We also provide a cost model to estimate the resource usage for a
given set of parameters as described in \autoref{sec:cost_model}. %

\textbf{Software Programmability.} %
Our hardware architecture is software-programmable at the granularity of
instructions as described in \autoref{sec:sw_stack}. %
This offers several advantages such as the ability to tailor block sizes and
dynamically skip bit positions for sparse or approximate computing. %

%  LocalWords:  textbf datapath

\subsection{Hardware Architecture Overview}
\label{sec:hw_architecture}

\begin{figure}[t]
  \centering
  \includegraphics[width=0.75\columnwidth]{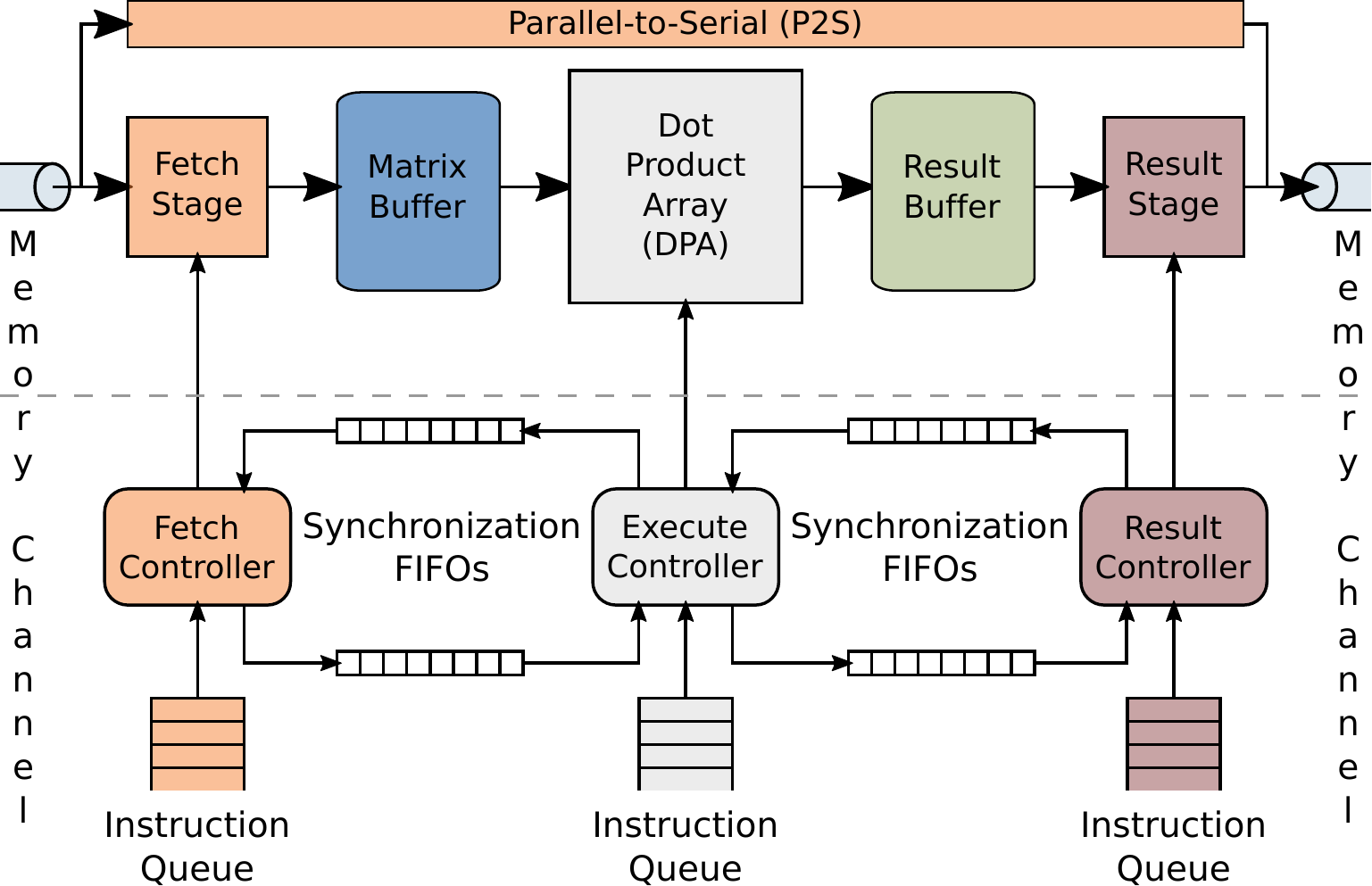}
  \caption{Overview of \OurScheme{}'s' hardware architecture.}
  \label{fig:hw_overview}
\end{figure}

% Overview of general structure of the accelerator
\autoref{fig:hw_overview} provides an overview of the \OurScheme{}
hardware. %
The architecture is organized into three pipeline stages: \textit{fetch},
\textit{execute}, and \textit{result}. %
Each stage communicates data to the next stage via shared on-chip memory
buffers. %
Inter-stage synchronization is achieved by blocking reads and writes to
synchronization FIFOs. %
All stage operations, including datapath control and synchronization, are
controlled by instructions, which are fetched from instruction queues and
executed in order. %
In addition to these stages, there is a Parallel-to-Serial (P2S) component for
data layout conversion (\autoref{sec:p2s_accelerator}), which is
incorporated into \OurScheme{} as an optional, standalone accelerator.

\begin{figure}[t]
  \centering
  \includegraphics[width=0.75\columnwidth]{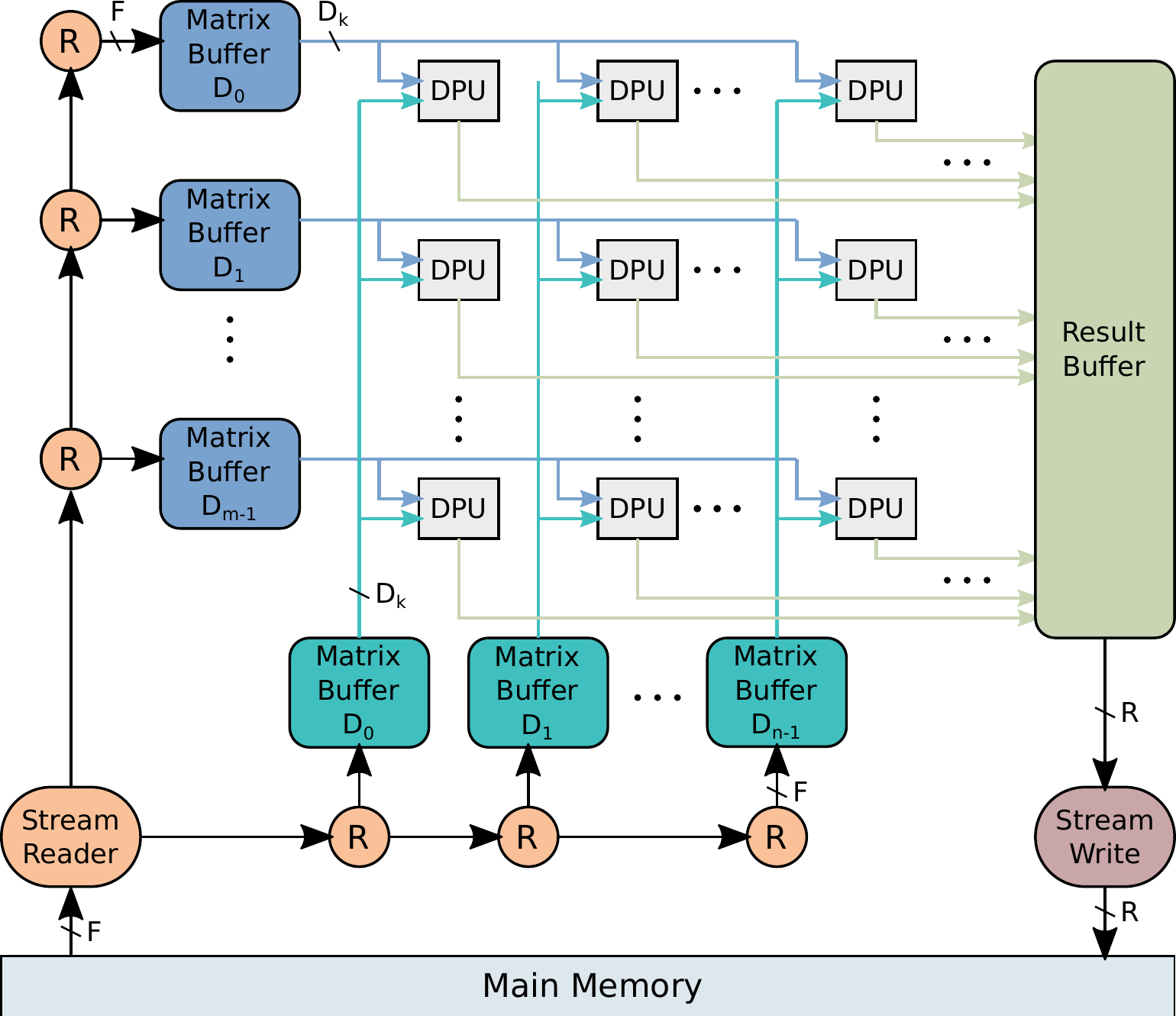}
  \caption{Key components of the \OurScheme{} datapath.}
  \label{fig:hw_datapath}
\end{figure}

% Overview of datapath, mention parametrization
The core of the hardware architecture is the bit-serial matrix-matrix
multiplication datapath illustrated in \autoref{fig:hw_datapath}. %
Accelerator performance and resource usage can be controlled by the
parameters specified in \autoref{tab:parameters}. %

\begin{table}
  \caption{Key \OurScheme{} hardware parameters.}
  \label{tab:parameters}
  \centering
  \begin{tabular}{cl}
    \toprule Symbol & Description \\
    \midrule
    $D_m, D_n$ & Number of DPUs in the DPA\\
    $D_k$ & DPU input bit width (popcount width) \\
    $B_m, B_n$ & Depth of input matrix buffers \\
    $B_r$ & Depth of result matrix buffer \\
    $A$ & Accumulator bitwidth \\
    $F$ & Main memory read channel bit width \\
    $R$ & Main memory write channel bit width \\
    $M$ & Maximum bit-parallel bitwidth for P2S \\
    \bottomrule
  \end{tabular}
\end{table}

% \TODO{add description of parallel-to-serial, either as part of fetch
% stage or as a standalone unit}

%\subsubsection{The Fetch Stage}
%\label{sec:fetch_stage}

% Overview
\textbf{The Fetch Stage} is responsible for reading matrix data from main memory
and populating the matrix buffers with data. %
Internally, the fetch stage contains a simple DMA engine and route generator
called a \textit{StreamReader}, as well as a linear array interconnect. %
% (highlighted in orange in \autoref{fig:hw_datapath}). %
% Operation
The StreamReader sends read requests to main memory and determines where read
responses are to be written, as specified by fetch instructions. %
% Interconnect
The read data and its destination form a packet that is carried through the
interconnect to the appropriate matrix buffer. %
The interconnect is bandwidth-matched to the main-memory read channel to
avoid any bottlenecks and ensure efficient use of off-chip bandwidth. %
% Synchronization, simple design, back pressure
The synchronization with the execute stage is ensured prior to fetching data,
which greatly simplifies the design of the interconnect as there is no
back pressure. %
% Scalable
The fetch stage can be scaled at design time to match the memory read
bandwidth ($F$) of a particular platform. %

%\TODO{Mention the design-time configurable matrix buffer size}

%\subsubsection{The Execute Stage}
%\label{sec:execute_stage}

% Overview
\textbf{The Execute Stage} is responsible for performing the matrix
multiplication on the data present in the matrix buffers. %
The core of the stage consists of an array of dot product units (DPUs), where
each DPU is fed with a design-time configurable number of bits ($D_k$) from the
left-hand-side and right-hand-side matrix buffers. %
The DPUs on the same row of the data processing array are fed with the same data
broadcasted by the left-hand-side matrix buffers. %
Similarly, the DPUs on the same column are fed with the same data broadcasted by
the right-hand-side matrix buffers (\autoref{fig:hw_datapath}). %
A single software controllable sequence generator is responsible for reading out
the appropriate data from the matrix buffers. %
The same generated sequence is used for both the left- and right-hand-side
matrix buffers but with different offsets. %
% Configuration
The execute stage can easily be scaled at design time by configuring the number
of rows ($D_M$) and columns ($D_N$) of DPUs. % in the DPA. %
Part of the contribution of this work is a version of the DPU that
is optimized for Xilinx FPGAs.
Both the original BISMO DPU and the improved version are described in further
detail in \autoref{sec:DPU}.

\textbf{The Result Stage} is responsible for writing the results generated by
the execute stage to main memory. %
The stage consists of a \textit{StreamWriter}, which contains a downsizer
(wide-in-narrow-out) to resize the array of results into the appropriate width
needed by the memory channel and a DMA engine with striding support to carry out
the actual memory write operations. %
The striding is needed to produce the result matrix one tile at a time. %
% \textcolor{blue}{HMS: the tiling has not been described so far so might be more
%   information than the reader can relate to.}
% , since results are produced on a a tile-by-tile basis. %
% Operation
When the execute stage has produced a new set of results, the accumulated
dot-products are written to the result buffer, from which the result stage writes
them to main memory. %
This enables the two stages to work independently and to overlap computation
and data transfer. %
% Scaling
The result stage can be scaled at design time to match the memory write
bandwidth ($R$) of a particular platform. %

% Once the execute stage indicating that a new set of results has been
% generated, the result stage can acquire this token and proceed to writing the
% result buffer's contents into main memory. %

% \TODO{Missing connection between accumulate register in the execute stage and
%   the result buffer. The accumulated result is transferred in a single cycle to
%   the result buffer, which is then written to memory over multiple cycles.}

\textbf{Parallel-to-Serial (P2S)} is an optional component that converts bit-parallel
matrices commonly used for CPUs into bit-serial ones as required by
\OurScheme{}.
The P2S does not communicate with the regular \OurScheme{} stages and is invoked as
a separate, standalone accelerator.
Its architecture is further described in \autoref{sec:p2s_accelerator}.

\subsection{The Dot Product Unit}
\label{sec:DPU}

The dot product unit (DPU) forms the core of the BISMO execute
stage. %
Each DPU performs a bit-serial dot-product operation between two
weighted binary vectors. %
Here, we start by describing the DPU of the original
BISMO~\cite{umuroglu+:FPL2018} and its shortcomings in terms of how it
maps to FPGAs (\autoref{sec:original_DPU}). %
Afterwards, we discuss a new DPU implementation with an FPGA-optimized
compressor (\autoref{sec:compressor}) and an improved datapath (\autoref{sec:new_DPU}). %

\subsubsection{Original BISMO DPU}
\label{sec:original_DPU}

\begin{figure}[t]
  \centering
  \includegraphics[width=0.75\columnwidth]{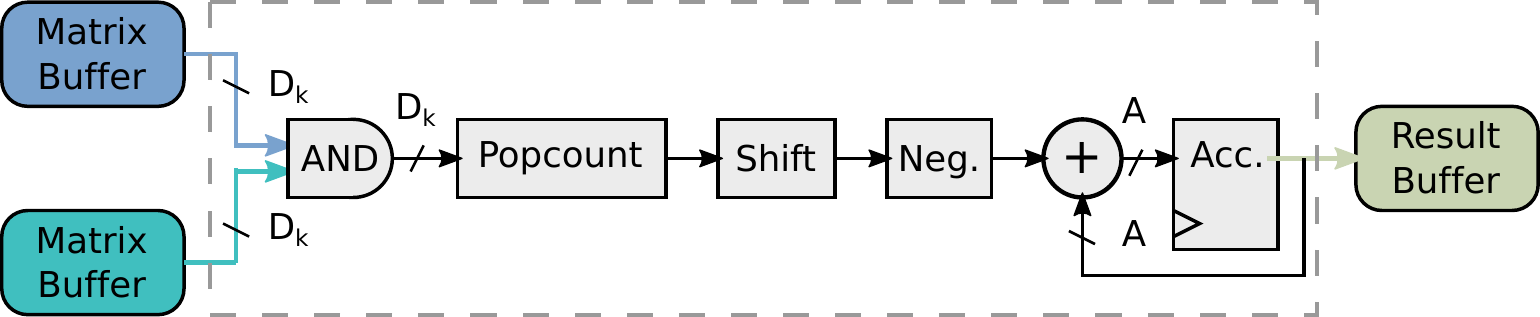}
  \caption{The original \OurScheme{} dot product unit (DPU).}
  \label{fig:hw_olddpu}
\end{figure}
% DPU operation
The original DPU~\cite{umuroglu+:FPL2018} can be seen in
\autoref{fig:hw_olddpu}. %
The DPU computes a partial result of the dot product between a row and
column of two bit-matrices, line 12 in
Algorithm~\ref{alg:bit-serial_matrix_multiplication}. %
The single-bit multiplications are performed by bitwise logic
\textsc{And} operations and the summation is a simple population count
(popcount) of the result. %
The weight in Algorithm~\ref{alg:bit-serial_matrix_multiplication} is
implemented by a left-shift unit and an optional negation, which are
controllable by software. %
The partial results are accumulated and stored in a register (Acc.) of
width $A$, which is typically 32 bits to avoid
overflows~\cite{umuroglu_jahre:CASES2017, umuroglu+:FPGA2017finn}. %
The shortcomings of this DPU architecture are twofold: %
\begin{enumerate}
\item The binary multiply and accumulate operation is implemented as a bitwise
  \textsc{And} followed by a popcount unit built as a tree of 6:3 popcount
  operators and adders. %
  Especially with large $D_k$, the popcount unit can require a large number of
  LUTs and many stages to pipeline the adder tree. %
\item The number of positions to left-shift the \textsc{And}-popcount
  result is supplied dynamically, which requires an expensive barrel
  shifter. %
\end{enumerate}

\subsubsection{Efficient \textsc{And}-Popcount for Xilinx FPGAs}
\label{sec:compressor}

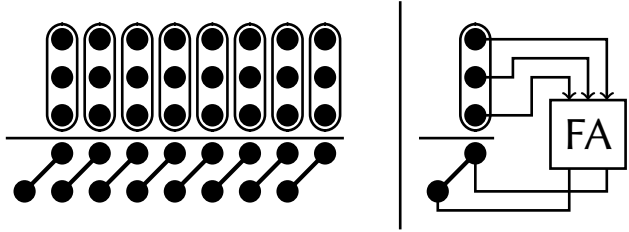
\begin{figure}
  \centerline{\begin{tikzpicture}[scale=.5,line width=1pt,style={font=\sffamily}]
  \foreach \x in {0,1,2,3,4,5,6,7,11} {
    \foreach \y in {-1,0,1,2}
      \filldraw(\x,\y) circle(.26);
    \filldraw(\x-1,-2) circle(.26);
    \draw[black,rounded corners=6](\x-.4,-0.4) rectangle(\x+.4,2.4);
    \draw [ultra thick] (\x,-1) -- (\x-1,-2);
  }
  \draw (-1.5,-0.6) -- (7.5,-0.6);
  \draw (9,-3) -- (9,3);
  \draw (9.5,-0.6) -- (11.5,-0.6);
  \draw (13,-1.4) rectangle (15, .4) node [midway] {\huge FA};
  \draw[->] (11,2) -- (14.5,2) -- (14.5,.4);
  \draw[->] (11,1) -- (12,1) -- (12,1.5) -- (14,1.5) -- (14,.4);
  \draw[->] (11,0) -- (12.5,0) -- (12.5,1) -- (13.5,1) -- (13.5,.4);
  \draw (13.5,-1.4) -- (13.5,-2.5) -- (10,-2.5) -- (10,-2);
  \draw (14.5,-1.4) -- (14.5,-2) -- (11,-2) -- (11,-1);
\end{tikzpicture}}
  \caption{Example matrix compression using carry-save-addition}
  \label{fig:csa}
\end{figure}

\looseness -1
The input to a popcount operation is a column of equally weighted bits,
which are to be summed up. While exhibiting an extreme aspect ratio of
$D_k\times1$, the input still forms a bit heap, which can be
reduced by standard matrix compression techniques. Step by step,
reshaped matrices are derived. They increase in width introducing
more and more higher weight bits but decrease in height while
always maintaining the numeric sum of the matrix rows. Only the final
summation into a single row representing the conventional binary result
requires an addition with a critical carry propagation. All preceding
compression steps can rely on parallel counters with bounded critical path
lengths that are independent from the matrix width.

For the general idea of carry-free bit heap compression,
refer to \autoref{fig:csa}. It shows a carry-save addition using regular
full adders operating in parallel to reshape a three-row input matrix into
a two-row output matrix with the same arithmetic sum. The customary
representation as a dot diagram abstracts the individual input and
output bits into plain dots. The numeric weight of each bit is determined
by its column, just as in the binary number system. In fact, when reading
each row as a binary number, the compression maintains the invariant that
the sum of the three input numbers equals the sum of the two output
numbers. The structural implementation of the compression is implied by
encircling the inputs to and connecting the outputs of each bit counter,
i.e., simple full adders in this case. Note that the carry outputs of
these full adders move up by a column as their numeric weights are two
times higher than that of the associated sum bits. Also, notice that the
combinational delay of this carry-save compression is a single full adder
irrespective of the actual width of the matrices.

More sophisticated parallel counters have been proposed and implemented
specifically targeting an efficient mapping to FPGA devices~\cite{parandeh:2011,
  kumm:2014, preusser2017generic, kumm:2018}. %
We leverage the open-source set of parallel counters and the associated generic
compressor implementation for Xilinx FPGAs proposed by
Preu{\ss}er~\cite{preusser2017generic}. %
It produces solutions optimized for our target FPGA architectures and integrates
easily into a regular synthesis flow. %
Its efficacious greedy scheduling of parallel counters avoids optimization
efforts that would be intolerable within a design cycle. %

The parallel counters used by the chosen generic compressor implementation are
mapped explicitly to concrete physical device primitives of the targeted Xilinx
devices. %
While this approach certainly enables highly optimized implementations, its
high degree of specialization also implies an inflexible operator interface.
It practically leaves no opportunities for the synthesis tool to optimize the
implementation within the context of the surrounding logic. %
In our particular case, we actually need a \textit{fused} \textsc{And}-popcount
operator. %
Optimizing the popcount alone isolates trivial 2-input \textsc{And}
gates at its inputs. These are, in the end, greatly underutilizing the
functional capabilities of the 6-input LUTs found on modern FPGA devices. %

\begin{figure}
  \begin{minipage}{.52\linewidth}\begin{center}
    \centerline{\scalebox{.7}{\usetikzlibrary{circuits.logic.US}
\begin{tikzpicture}[circuit logic US,scale=.8,line width=1.4pt,style={font=\sffamily}]
% AND Gates and FA
\node at(1.75,1.5) {\Large $\vdots$};
\foreach \y in {0,1,2} {
	\node[and gate,minimum height=36] (and\y) at (0.2,-1.5*\y) {};
	\draw ([yshift= 2]and\y.input 1) -- +(-.8,0) node[left] {$a_{3i+\y}$};
	\draw ([yshift=-2]and\y.input 2) -- +(-.8,0) node[left] {$b_{3i+\y}$};
	\draw (and\y.output) -- (2.5,-1.5*\y) node[lightgray] (p\y) [fill,circle,minimum height=14] {}
		 -- +(1,0) |- (4.1,-1-.5*\y);
}
\draw[lightgray,rounded corners=12]($(p0)+(.5,.5)$) rectangle($(p2)-(.5,.5)$);
\draw (4.1,-2.3) rectangle +(1.8,1.6) node[midway] {\Huge FA};
\draw[thin] (-1,.9) rectangle +(7.2, -4.8) node[below right,xshift=-16] {$2\times$ 6-LUT};
\draw[thin,gray,dotted] (-.9,.8) rectangle +(7.2, -4.8);
\node at(1.75,-4.2) {\Large $\vdots$};

% Reduced Output
\draw[line width=2pt](0,-5) -- (3.5,-5);
\node at(1.75,-5.4) {\Large $\vdots$};
\foreach \i in {0,1} {
	\draw (5.3-.6*\i,-2.3) -- +(0,-3.8+.25*\i) -| (2.5-1.5*\i,-6.6) node[lightgray] (p) [fill,circle,minimum height=14] {};
}
\draw[lightgray] (2.5,-6.6) -- (1,-6.6);
\node at(1.75,-7.2) {\Large $\vdots$};
\end{tikzpicture}}}
    \caption{Fused pre-compression of three bit products}
    \label{fig:precompress}

    \vspace*{\baselineskip}

    \captionof{table}{Bit counter statistics across operator sizes}
    \label{tab:counter_stats}
    \begin{tabular}[c]{rrrrr}\toprule
      $N$ & \textbf{(2,5:*]} & \textbf{(6:3]} & \textbf{(3:1]} & \textbf{Slice}\\\midrule
        32 &  3 &  1 &   &    \\
        64 &  4 &  3 & 2 &  1 \\
       128 &  8 &  7 & 5 &  3 \\
       256 & 17 & 13 & 3 &  9 \\
       512 & 38 & 25 & 4 & 19 \\
      1024 & 79 & 51 & 6 & 38 \\\bottomrule
    \end{tabular}
  \end{center}\end{minipage}\hfill
  \begin{minipage}{.46\linewidth}
    \centerline{\resizebox{!}{.5\textheight}{\usetikzlibrary{circuits.logic.US}
\begin{tikzpicture}[circuit logic US,scale=1,line width=1.4pt,style={font=\sffamily}]

% Stage #1
\foreach \i in {-1,...,9} {
	\draw (0,\i) node[fill,circle,minimum height=14] {} -- (1, \i) node[fill,circle,minimum height=14] {};
}
\draw (0.5,-1) node{/};
\draw[black,rounded corners=12](-.4,1.6)rectangle(.4,7.4);
\draw[black,rounded corners=12](-.4,-.4) -- ++(1.8,0) -- ++(0,4.8) -- ++(-.8,0) -- ++(0,-3) -- ++(-1,0) -- cycle;
\draw[black,rounded corners=12](-.4,9.4) -- ++(1.8,0) -- ++(0,-4.8) -- ++(-.8,0) -- ++(0,3) -- ++(-1,0) -- cycle;
\draw(-3.5,-1.5) -- ++(5,0);

% Stage #2
\foreach \i in {-2,-4} {
	\draw (-1,\i) node[fill,circle,minimum height=14] {}
		-- (1, \i) node[fill,circle,minimum height=14] {};
	\draw (0,\i) node[fill,circle,minimum height=14] {}
		-- (0, \i-1) node[fill,circle,minimum height=14] {};
}
\draw (-2,-6) node[fill,circle,minimum height=14] {}
	-- ++(1, 0) node[fill,circle,minimum height=14] {}
	-- ++(1, 0) node[fill,circle,minimum height=14] {};
\draw (0,-7) node[fill,circle,minimum height=14] {};
\draw (1,-7) node[fill,circle,minimum height=14] {};
\draw[black,rounded corners=12](-1.4,-1.6) -- ++(1.8,0) -- ++(0,-4.8) -- ++(-.8,0) -- ++(0,2) -- ++(-1,0) -- cycle;
\draw(-3.5,-7.5) -- ++(5,0);

% Stage #3
\draw (-2,-8) node[fill,circle,minimum height=14] {}
	-- (0, -8) node[fill,circle,minimum height=14] {};
\draw (-1,-8) node[fill,circle,minimum height=14] {}
	-- (-1, -9) node[fill,circle,minimum height=14] {};
\foreach \i in {-8,...,-10} \draw (1,\i) node[fill,circle,minimum height=14] {};
\foreach \i in {-2,...,0} \draw (\i,-10) node[fill,circle,minimum height=14] {};

% CPA
\draw(-3.4,-11.4)rectangle++(.8,.8) node[midway] {HA};
\draw(-2.4,-11.4)rectangle++(.8,.8) node[midway] {TE};
\draw(-1.4,-11.4)rectangle++(.8,.8) node[midway] {TE};
\draw(-.4,-11.4)rectangle++(.8,.8) node[midway] {FA};
\draw(.6,-11.4)rectangle++(.8,.8) node[midway] {FA};
\draw (1.4,-11.2) -- +(.4,0) |- (1,-10);
\foreach \i in {-3,-2} \draw (\i+.4,-10.8) -- +(.2,0);
\foreach \i in {-3,...,0} \draw (\i+.4,-11.2) -- +(.2,0);
\foreach \i in {-3,...,1} \draw (\i,-11.4) -- +(0,-0.4);
\draw (-3.4,-11) -| +(-.4,-.8);
%\draw (-2.4,-12.2) -| +(-.4,-.6);
\end{tikzpicture}}}
    \caption{Continued compression of 32-bit popcount}
    \label{fig:compress32}
  \end{minipage}
\end{figure}

In order to eliminate this interfacing inefficiency, we designed a physically
fused operator implementation by preceding the generic compressor with an
equally rigorously optimized pre-compression. %
Instead of computing individual bit products, they are combined into groups of
three whose computations are absorbed into the equivalent of a full-adder
compression. %
Note that all these groups can be pre-compressed independently and in
parallel. %
The computation implementing this functionality is depicted in
\autoref{fig:precompress}. It can be mapped directly to two 6-input LUTs. %
It is worth noting that this pre-compression favorably changes the geometry of
the bit heap input to the generic compressor. %
Instead of feeding a $D_k\times1$ matrix, the pre-compression already reduces
this height to $\left\lceil D_k/3\right\rceil$ while spreading the input across
two columns. %

The structure of the complete summation process of a 32-bit popcount
operation is illustrated in \autoref{fig:compress32}. Following the convention to
encircle the inputs and to connect the outputs of a counter primitive,
it shows the summation structure generated
by the algorithm proposed by Preu{\ss}er~\cite{preusser2017generic}. It comprises
two compression steps with parallel bit counters and a final carry-propagating
ternary addition. Identifying the bit counters by the individual heights of their
input columns from left to right, a pair of $(5,2)$-counters and a $(6)$-counter
accomplish the first parallel compression step. This is only followed by one
other small compression through a single $(5,2)$-counter prior to the
carry-propagating summation. In this particular case, only the third column of
the compression result is high enough to introduce the second,
only locally forwarded carry signal that is typical for a ternary addition.
The carry propagation chain is terminated by a final half adder.

The first compression step is dominated by $(5,2)$-counters for all
larger operator sizes. The pre-compressed two-column input is too narrow for
slice-based counters leaving the $(5,2)$-counters as the most economic choice.
The bits their application leaves, predominantly in the second column, are mostly
handled by $(6)$-counters just as shown in the 32-bit example. Full adders
would take care of fewer leftover bits. As larger operator implementations
reach wider intermediate bit heap geometries before the final addition, they
will also utilize slice-based counters in these later compression steps.
These counters leverage the carry chain to combine four LUTs of a slice to obtain
counter primitives optimized for the target device architecture . An overview of
the use of the different
counters in our designs is given by Tab.\,\ref{tab:counter_stats}. %\autoref does not honor \captionof

We employ a fully pipelined compressor with register stages separating all
compression steps to optimize the operating frequency of the
\textsc{And}-popcount reduction. %
It is worth to mention that we had to replace the trivial behavioral register
description by an explicit instantiation of \texttt{FDRE} register primitives in
order to avoid excessively growing synthesis times for larger operators. %
It appeared that the Xilinx Vivado synthesis engine~\cite{vivado_v2017.4} had a
hard time or was trying too hard to optimize the many interfaces between behavioral code and the netlists of primitives generated for the compression steps.

\subsubsection{Improved DPU}
\label{sec:new_DPU}

\begin{figure}[t]
  \centering
  \includegraphics[width=0.75\columnwidth]{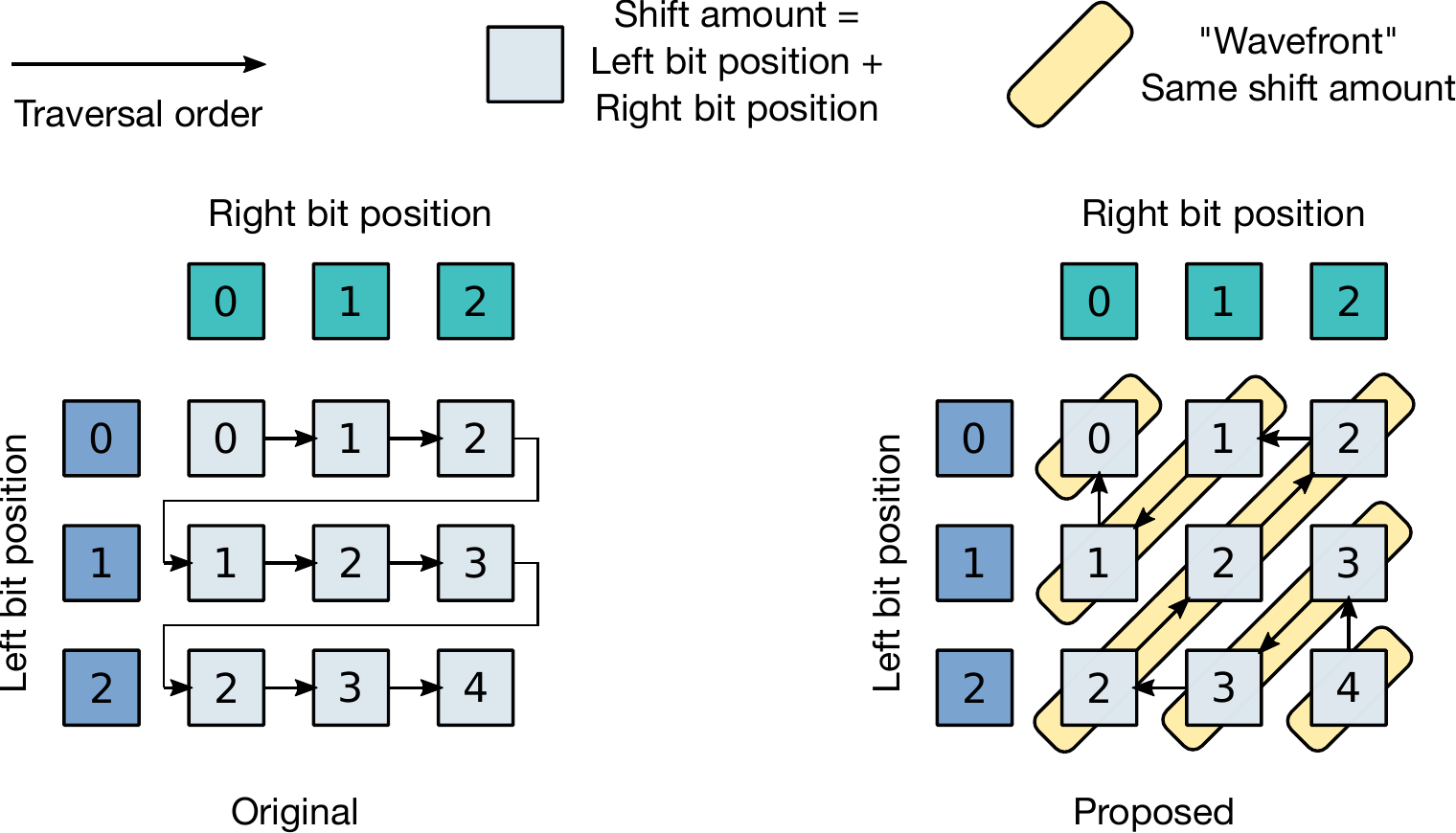}
  \caption{Original and proposed bit position traversal order.}
  \label{fig:hw_bitorder}
\end{figure}

\begin{figure}[ht]
  \centering
  \includegraphics[width=0.75\columnwidth]{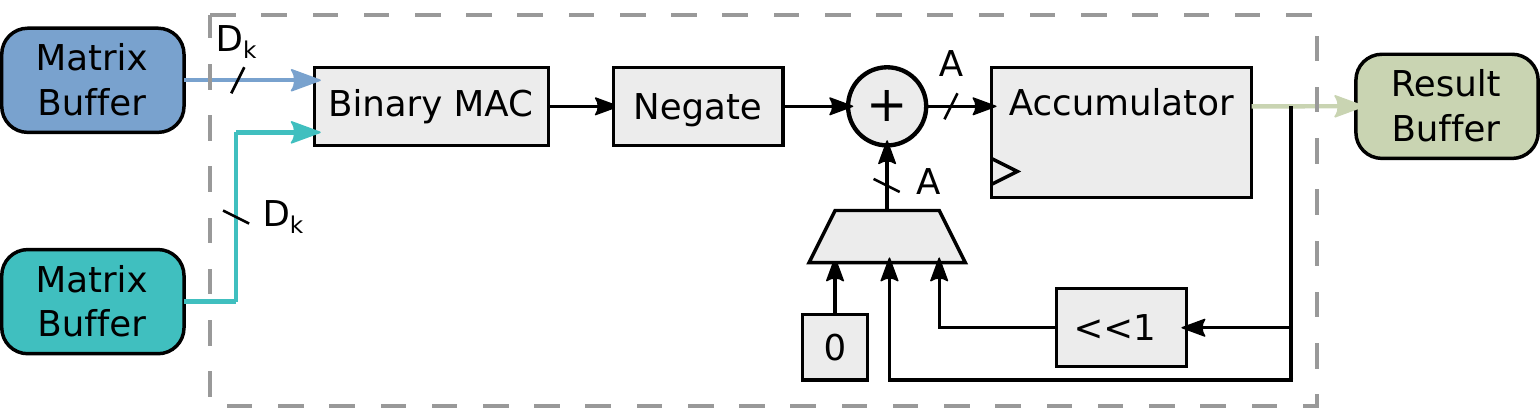}
  \caption{The optimized dot product unit (DPU).}
  \label{fig:hw_newdpu}
\end{figure}

The barrel shifter in the original BISMO DPU is needed to account for the
differences in weight between the accumulator and the contribution. %
This difference depends on the order in which the bits of the $L$ and $R$ matrices
are traversed. %
First, we note that the loop nest in
Algorithm~\ref{alg:bit-serial_matrix_multiplication} is affine, and the $L$ and $R$
bit positions (variables $i$ and $j$) can be traversed in any order as long as
the correct weight is applied. %
Based on this observation, we propose to traverse the bit positions as shown in
\autoref{fig:hw_bitorder}. %
Here, the sum of $L$ and $R$ bit positions constitute \textit{wavefronts}, where
each wavefront has a left-shift value that is one less than the previous one. %
Using this schedule, instead of left-shifting the current contribution by a
variable amount with a barrel shifter, we can left-shift the previous
accumulator by either one position (if changing wavefronts) or use it as-is,
before summing the accumulator and the contribution. %
The optional negation is still applied to the current contribution, when needed
for bit position combinations that yield a negative result. %
Combined with the new \textsc{And}-popcount unit (\autoref{sec:compressor}),
this yields the improved DPU design illustrated in \autoref{fig:hw_newdpu},
where the barrel shifter is replaced with a constant one-left-shift and a
multiplexer. %
Barring accumulator overflows, our new DPU is able to handle any input
precision, whereas the original DPU was limited by the maximum left-shift
supported by the barrel shifter.

%  LocalWords:  includegraphics columnwidth vspace textit datapath textbf
%  LocalWords:  toprule midrule popcount bottomrule subsubsection dpu
%  LocalWords:  textcolor downsizer umuroglu fig:hw_olddpu textsc
%  LocalWords:  alg:bit-serial_matrix_multiplication fig:csa Preu
%  LocalWords:  umuroglu_jahre:CASES2017 kumm:2014 kumm:2018 lceil
%  LocalWords:  preusser2017generic fig:precompress rceil texttt
%  LocalWords:  Vivado netlists fig:hw_bitorder fig:hw_newdpu rrrrr
%  LocalWords:  parandeh:2011 bitwidth minipage linewidth scalebox
%  LocalWords:  baselineskip captionof hfill resizebox textheight

\subsection{Bit-Parallel to Bit-Serial Matrix Transformation}
\label{sec:p2s_accelerator}

\begin{figure}[t]
  \centering
  \includegraphics[width=0.75\columnwidth]{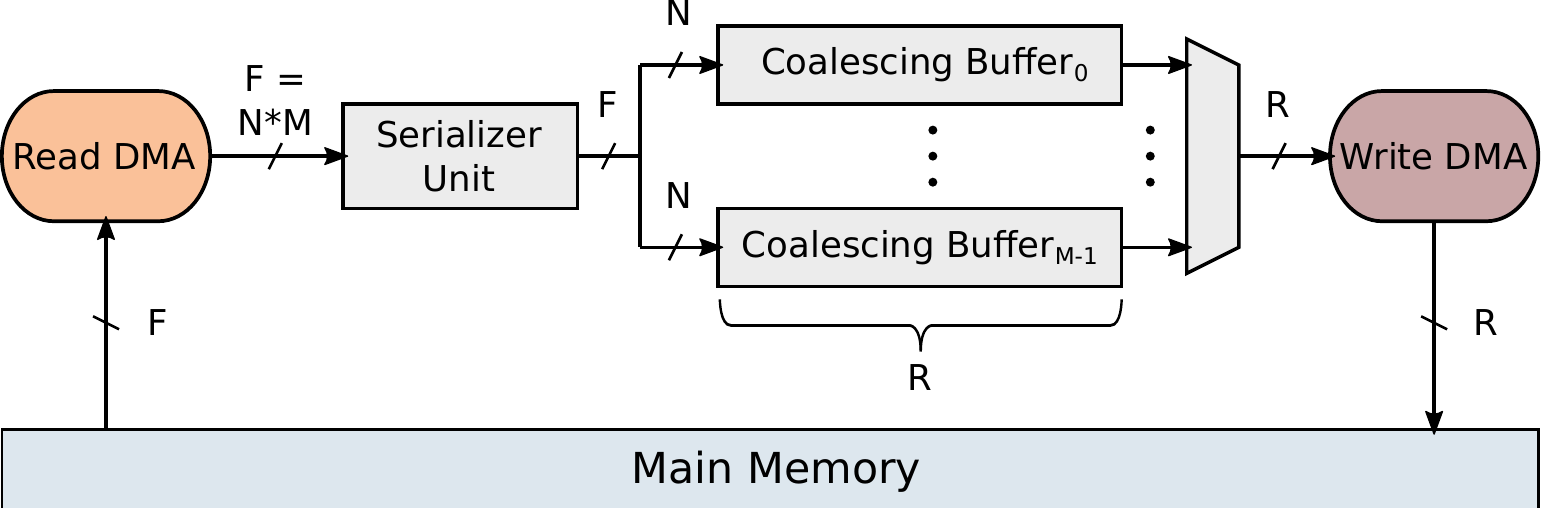}
  \caption{High-level view of the P2S accelerator.}
  \label{fig:p2s_high_level}
\end{figure}

As described in \autoref{sec:bsdatalayout}, BISMO assumes that the input
matrices are present in main memory, using a bit-serial data layout. %
However, due to the bit-parallel nature of the arithmetic in general-purpose
CPUs, matrices are almost always stored using a bit-parallel data layout in
practice. %
Furthermore, as CPUs typically offer 8-bits as the smallest native data type,
matrices that require fewer bits are also stored using 8-bit data types. %
Conversion from bit-parallel to bit-serial can be a costly operation, whose cost
must be taken into account as part of the accelerator performance. %

\begin{figure}[t]
  \centering
  \includegraphics[width=1\columnwidth]{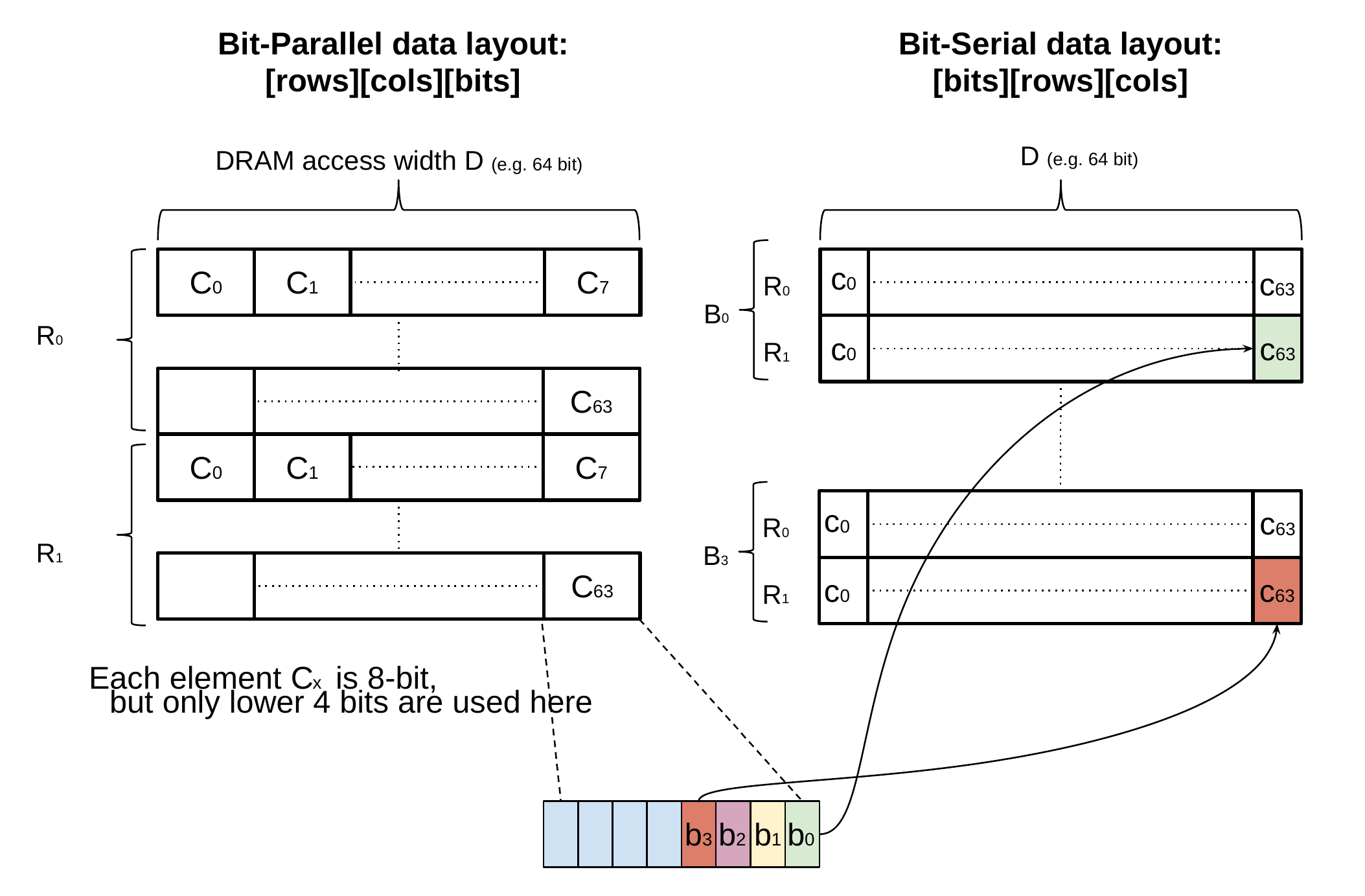}
  \caption{An example of a memory layout for an example matrix of 2x64 of 4 bits.}
  \label{fig:p2s_data_layout}
\end{figure}

To address this problem for BISMO, we enhance it with a stand-alone
parallel-to-serial (P2S) accelerator. %
The accelerator, illustrated in \autoref{fig:p2s_high_level}, is a data-layout
transformer with run-time configurable precision. %
The P2S retrieves a bit-parallel matrix (left-hand side in
\autoref{fig:p2s_data_layout}), transforms it into a bit-serial matrix
(right-hand side), and writes it back to main memory. %
The P2S \textit{read DMA} sequentially fetches the column elements constituting
a row of the bit-parallel matrix from main memory and feeds these into the
\textit{serializer unit}. %
The individual bits of each column element are split up across as many
\textit{coalescing buffers} as the bit-precision of the parallel matrix. %
This is repeated for all the rows of the input matrix. %
% Runtime configurable parameters
The bit precision and number of rows and columns of the parallel matrix are
runtime configurable. %
% Synthesis configurable parameters
The total number of coalescing buffers defines the maximum supported precision
$M$ of the bit-parallel input matrix and is specified at synthesis time. %
The coalescing buffer size is given by the bitwidth of the write bus, which is
also specified at synthesis. %
We assume that the bitwidths of the read and write buses are given by the
\OurScheme{} parameters $F$ and $R$ specified in \autoref{tab:parameters}.

% \textcolor{blue}{I removed the mentioning of M, R and F as these are not well
%   defined and I cannot even find R and F in the figures.}

\autoref{fig:p2s_data_layout} illustrates an example of a 4-bit parallel matrix
of size 2x64 where each column element has been padded to eight bits and 64-bit
read and write data buses are used. %
Eight column elements (total 64 bits) of the bit-parallel matrix are fetched on
each memory access. %
The four most significant bits of each 8-bit column element are padding and are
discarded (i.e., the actual precision is specified to be four bits by the P2S
instruction at runtime). %
The remaining four bits are split across four different coalescing buffers, one
for each bit weight (B$_0$-B$_3$). %
The column index within the row dictates the bit position written in the
coalescing buffers. %
As shown in the example, the final column (C$_{63}$) of the second row (R$_1$)
is written to the last bit position (c$_{63}$) of the coalescing buffers that
are allocated for the row (B$_0$-B$_3$). %
If the row of the bit-parallel matrix contains more columns than bit positions
in the coalescing buffer, then the P2S kernel stalls to write back the
coalescing buffers to main memory before continuing the transformation of the
remaining columns. %
The allocated coalescing buffers are also written back to main memory when a new
row is encountered in the bit-parallel matrix (e.g., R$_1$). %

To simplify the implementation, the number of columns of the bit-parallel matrix
has to be a multiple of the coalescing buffer bit-width ($R$). %
This requires some input matrices to be padded but greatly simplifies the write
back of the bit-serial matrix. %
This ensures that the coalescing buffers are completely filled and can be
written back to memory without requiring the data to be realigned. %
The binary matrices are stored consecutively, i.e., all the rows of binary
matrix B$_0$ are stored together which are then followed by B$_1$ and so on. %
This requires the coalescing buffers to be written back in a strided fashion
with B$_0$-R$_0$ being written together with B$_1$-R$_0$ and a stride equal to
the size of a complete binary matrix. %

\subsection{Cost Model}
\label{sec:cost_model}

For any parametrizable overlay architecture, it is beneficial to provide a model
of how the FPGA resource usage relates to its configuration parameters. %
This enables a quick performance estimation when scaling to other devices. %

\subsubsection{LUT cost}
\label{sec:costmodel_LUT}
We propose the following equations to model the LUT usage of a \OurScheme{}
instance:
\newcommand{\lutcost}[1] {\mathrm{LUT}_{\mathrm{#1}}}
\newcommand{\mconst}[1] {\alpha_{\mathrm{#1}}}
\newcommand{\aconst}[1] {\beta_{\mathrm{#1}}}
\newcommand{\bramcost}[1] {\mathrm{BRAM}_{\mathrm{#1}}}

\begin{subequations}
\vspace{-2mm}
\begin{equation} \label{eqn:lutcost_total}
    \lutcost{total} = \lutcost{base} + \lutcost{array}
\end{equation}
\vspace{-5mm}
\begin{equation} \label{eqn:lutcost_array}
  \lutcost{array} = D_m \cdot D_n \cdot (\lutcost{DPU} + \lutcost{res})
\end{equation}
\vspace{-5mm}
\begin{equation} \label{eqn:lutcost_dpu}
  \lutcost{DPU} = \mconst{DPU} \cdot D_k + \aconst{DPU}
\end{equation}
\end{subequations}

\autoref{eqn:lutcost_total} breaks the total cost into $\lutcost{base}$,
which covers the DPA size-independent LUT usage such as the DMA engines, P2S
and other fixed platform infrastructure, and $\lutcost{array}$ which covers
the DPA size-dependent part.
In turn, \autoref{eqn:lutcost_array} further breaks down $\lutcost{array}$
into LUT cost for the DPU and for result generation, multiplied by the array
size.
Finally, we model $\lutcost{DPU}$ as a linear function of the popcount
width $D_k$ in \autoref{eqn:lutcost_dpu}, and $\lutcost{res}$ as a constant.
The constants $\mconst{DPU}, \aconst{DPU}, \lutcost{base}$ and $\lutcost{res}$ are
determined empirically in \autoref{sec:synthesis}.

\subsubsection{BRAM cost}
\label{sec:costmodel_BRAM}
Assuming dual-port $36 \times 1024$-bit Xilinx BRAMs, we model their usage as: %
% follows: %

\begin{subequations}
  \begin{equation} \label{eqn:bramcost_total}
      \bramcost{total} = \bramcost{base} + \bramcost{array}
  \end{equation}
\begin{equation} \label{eqn:bramcost_array}
    \bramcost{array} = \ceil*{\frac{D_k}{32}}  \cdot \left (D_m \cdot \ceil*{\frac{B_m}{1024}} + D_n \cdot \ceil*{\frac{B_n}{1024}} \right )
\end{equation}
\end{subequations}

In \autoref{eqn:bramcost_total}, $\bramcost{base}$ refers to the BRAMs used for
DPA-size independent infrastructure, such as DMA buffers and instruction
queues. %
$\bramcost{array}$ is the cost for the input matrix buffers. %
We use 32 of the native 36-bit width due to constraints from the fetch stage,
since DRAM buses are typically power-of-two-wide and we require BRAM read/write
widths to be an integer multiple of each other. %
We assume that the result matrix buffer consists of small LUTRAM buffers, and cover
their cost in \autoref{eqn:lutcost_array}. %

% \subsubsection{P2S LUT cost}
% \label{sec:P2S_cost_model}
%
% We model the cost of the standalone P2S accelerator separate from the
% \OurScheme{} cost. %
% The primary contributor to P2S resources is the number coalescing buffers, which
% is determined by the maximum bit-parallel precision $M$. %
% Thus, we model $\lutcost{P2S}$ as a linear function of $M$ as in
% \autoref{eqn:lutcost_p2s}, where $\mconst{P2S}$ and $\aconst{P2S}$ are
% determined empirically in \autoref{sec:synthesis}. %
%
% \begin{equation} \label{eqn:lutcost_p2s}
%   \lutcost{P2S} = \mconst{P2S} \cdot M + \aconst{P2S}
% \end{equation}

%  LocalWords:  subsubsection sec:costmodel_LUT newcommand lutcost mathrm cdot
%  LocalWords:  mathrm mconst bramcost subequations vspace eqn:lutcost_total
%  LocalWords:  eqn:lutcost_array cdot eqn:lutcost_dpu popcount frac frac frac
%  LocalWords:  sec:costmodel_BRAM eqn:bramcost_total eqn:bramcost_array
%  LocalWords:  eqn:lutcost_p2s

\subsection{Programming \OurScheme{}}
\label{sec:sw_stack}
\newcommand{\ISA}{Instruction}
\newcommand{\channelID}{Channel ID}
\newcommand{\schannelID}{channel ID}
\newcommand{\PutToken}{\texttt{Signal}}
\newcommand{\GetToken}{\texttt{Wait}}
\newcommand{\RunInst}{\texttt{Run}}
\newcommand{\RunFetch}{\texttt{RunFetch}}
\newcommand{\RunExecute}{\texttt{RunExecute}}
\newcommand{\RunResult}{\texttt{RunResult}}
\newcommand{\RunPtS}{\texttt{RunP2S}}
\newcommand{\putverb}{-}
\newcommand{\getverb}{-}

% \TODO{*Microcode Generator sounds like a hardware structure, but I
% assume that this is the software running on the core. Should be
% clarified.}

% \TODO{*Microcode has a certain association for computer architects,
% which is probably not correct for the context of BISMO instructions}

\OurScheme{} provides programmability through the use of instructions that
control each of the pipeline stages and the P2S. %
Taking into account the dimensions of the input matrices and the data layout in
memory, it is possible for a programmer to perform scheduling in various ways. %
The capabilities facilitated by these instructions and their usage are
illustrated in this section. %

\subsubsection{\ISA{}s}

\looseness -1
There are three types of instructions per pipeline stage in \OurScheme{}, namely
\GetToken{}, \PutToken{} and \RunInst{}. %
The P2S is treated as a separate accelerator synchronized at a coarser level,
and only has \RunPtS{}.
\autoref{tab:ins_desc} provides a summary of these instructions with the usage
described as follows: %
%The instructions \GetToken{} and \PutToken{} are used for synchronization between two
%pipeline stages and enable efficient data communication between them. %
% In order to perform the synchronization in communication of data between two
% pipeline stages, the instructions \textit{\GetToken{}} and
% \textit{\PutToken{}} are used.%
%The \RunInst{} instruction is used for setting input parameters and
%performing the particular functionality of a pipeline stage. %
% For configuration of input parameters and performing the particular
% functionality of a pipeline stage the \textit{\RunInst{}} instruction is used.%

\begin{table}
  \caption{\OurScheme{}'s Instruction Summary}
  \label{tab:ins_desc}
  \centering
% BEGIN RECEIVE ORGTBL
\begin{tabular}{ll}
\hline
Instruction type & Fields \\
\hline
\GetToken{} \& \PutToken{} & Associated FIFO: \\
 & ~~ Fetch stage: Execute \\
 & ~~ Execute stage: Fetch or Result \\
 & ~~ Result stage: Execute \\
\hline
\RunFetch{} & Source (main memory) parameters: \\
 & ~~ Base address \\
 & ~~ Block size (bytes) \\
 & ~~ Block offset (bytes) \\
 & ~~ Number of blocks to fetch \\
 & Destination (matrix buffer) parameters: \\
 & ~~ Matrix buffer offset \\
 & ~~ Starting matrix buffer \\
 & ~~ Range of matrix buffers \\
 & ~~ Consecutive words per matrix buffer \\
\hline
\RunExecute{} & Matrix buffer offset \\
 & Dot product length \\
 & Negate contribution mode\\
 & Accumulator shift mode \\
\hline
\RunResult{} & Result base address in main memory \\
 & Address offset \\
\hline
\RunPtS{} & Bit-parallel base address in main memory \\
 & Bit-serial base address in main memory \\
 & Number of rows and columns \\
 & Actual precision \\
\hline
\end{tabular}
% END RECEIVE ORGTBL
\end{table}

\textbf{The Synchronization Instructions}
are used for synchronization between two different pipeline stages. %
%Synchronization between two different stages are handled by the \GetToken{} and
%\PutToken{} instructions. %
The \PutToken{} instruction issues a token to the associated synchronization
FIFO, while the \GetToken{} instruction blocks on the associated synchronization
FIFO until it receives a token. %
For both the fetch and result stage, the only associated synchronization FIFO is
their respective FIFO for the execute stage. %
The execute stage has consequently two associated FIFOs for synchronization with
either the fetch or the result stage. %
The tokens do not convey any information and a programmer is free to decide what
each synchronization represents, e.g., that a particular matrix buffer is now
full or empty. %
We note that the P2S is treated as a separate accelerator synchronized at a
coarser level, and cannot be the source or destination for any synchronization
instructions.

% For a certain pipeline stage to wait for another to finish its function to
% obtain its data, the \GetToken{} instruction is issued.  The instruction queue
% will wait until the other pipeline stage has issued a token through
% \PutToken{} to the same FIFO, after which the \GetToken{} instruction can
% obtain that token to let the succeeding instruction to be executed.
% For both the fetch and result stage in \OurScheme{}, synchronization
% happens only with the execute stage.
% However in the case of the execute stage, it can synchronize with either the fetch or
% the result stages.
% Thus the synchronization instructions of the execute stage are modified to be able to choose
% between the fetch or result stage for synchronization.

% \TODO{Design-time configuration is mentioned regarding the HW parameters that
% can be controlled when generating a \OurScheme{}, \autoref{tab:parameters}. It
% would be good to use a different term when mentioning the software.}

\textbf{The Run Instructions}
are used to carry out the particular function of a pipeline stage. %

The \RunFetch{} instruction specifies from where in main memory to read data and
the destination matrix buffers to store the read data. %
% Main memory
The parameters with regard to main memory are: %
\textit{i)}~the base address from where the fetch should begin, %
\textit{ii)}~the size of the contiguous block to be fetched, %
\textit{iii)}~the offset between such blocks (providing strided accesses), and %
\textit{(iv)}~the number of blocks to be fetched. %
% Matrix buffers
The parameters with regard to matrix buffers are: %
\textit{i)}~the buffer offset at which to start writing data, %
\textit{ii)}~the matrix buffer to begin writing to (all buffers are enumerated
from zero to $D_m \cdot D_n - 1$), %
\textit{iii)}~the range of matrix buffers to be written (number of consecutive
buffers), and %
\textit{iv)}~the number of consecutive words to be written in each matrix buffer
before switching to the next. %
These set of parameters enable consecutive data blocks to be placed in one
matrix buffer before moving to the next or to place the blocks in a cyclic
fashion across a range of buffers. %

% To configure the destination matrix buffer locations are the matrix buffer tags
% (numeric tags arranged in ascending order starting from 0), matrix buffer range
% which corresponds to consecutive matrix buffers to which the fetched data should
% be placed, and the number of consecutive words that should be placed in each
% matrix buffer before switching to the next. %
% These set of parameters allows for consecutive data blocks to be placed in one
% matrix buffer before moving to the next or to place them in a cyclic fashion. %

The \RunExecute{} instruction specifies %
% Buffer offset
the matrix buffer offset from where to begin reading data, how many buffer
addresses will be read,
% Negation and accumulation modes
whether to negate the current contribution, and whether to accumulate with a
zero, the accumulator register, or the accumulator register left-shifted by one.

% The parameters to configure the execute stage are the number of tiles
% corresponding to a valid matrix multiplication result the bit-shift associated
% with this multiplication, a value to negate the resulting multiplication if
% needed, a value to clear any previously accumulated values (if the computed
% matrix tile is new), offsets from the starting address for the matrix buffers
% associated with the left hand side and the right hand side matrix to begin the
% computation. %

The \RunResult{} instruction specifies the base address of the result matrix
stored in main memory and an offset to which the current results are to be
written. %

% For the result stage the configuration parameters are the base address in the
% main memory and an offset so that the value corresponding to a particular matrix
% tile is written in the correct location. %

%In order for the Fetch stage to populate matrix buffers with data from the main memory,
%a \GetToken{} instruction will be issued to the FIFO channel connecting the execute stage
%to the Fetch Stage. The instructions sequence succeeding the \GetToken{} instruction will
%wait until a token is obtained from the execute fetch FIFO
%Each run of the fetch stage corresponds to gathering one or more blocks of data
%from the main memory, and distributing this data into the matrix buffers.
%Once a synchronization token is received from the execute stage indicating that
%a new set of results has been generated, the result stage can acquire this token
%and proceed to writing the result buffer's contents into main memory. %
%SW is able to keep one of the matrices always on-chip if small enough (e.g., for
%QNNs), flexible/don't have to assume input matrices always live in main memory.

Finally, the \RunPtS{} instruction specifies a source matrix expressed by its
base address in main memory, spatial dimensions (rows and columns) and actual
bit-parallel precision, i.e., how many bits starting from the least significant
bit should be converted, as well as, a main memory address where the resulting
bit-serial matrices are to be written. %

\newcommand{\mult}{$\cdot$}
\subsubsection{Instruction Scheduling}
\label{sec:schedexample}
\begin{figure}[t]
  \centering
  \includegraphics[width=0.75\columnwidth]{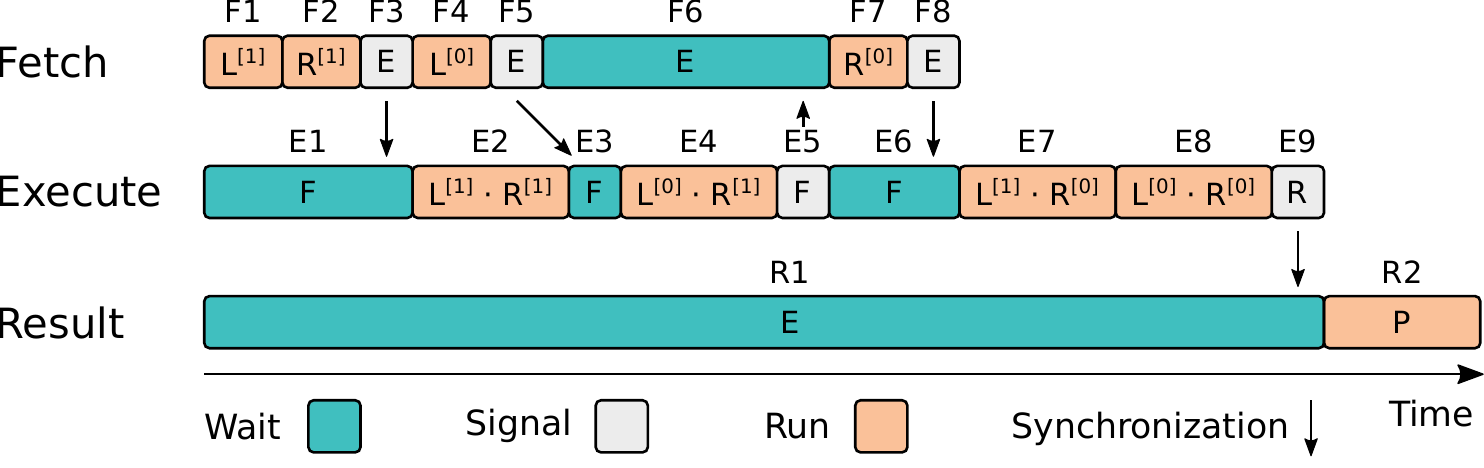}
  \caption{Timeline of the example schedule shown in \autoref{tab:instruction_queues}.}
  \label{fig:timeline}
\end{figure}

% It is important to consider the data layout and arrange the sequence of
% instructions for each pipeline stage to ensure correctness of the result when
% using \OurScheme{}. This can be illustrated with the following example.
%
%The \OurScheme{} instructions enable the possibility to tailor the
%matrix multiplication
%computation to the input matrix characteristics, e.g., by taking their
%dimensions into account. %
Using conventional block matrix multiplication algorithms that were previously
applied to FPGA matrix multiplication accelerators~\cite{matam2013energy},
\OurScheme{} can process matrices of any dimension.  \autoref{fig:timeline}
shows one possible schedule for the matrix multiplication example in
\autoref{fig:bit-serial_example}. %
% Assumption
Here, the DPA is assumed to be as large as the input matrices for simplicity. %
The computation would otherwise have to be divided into separate tiles resulting
in many more instructions. %
Furthermore, it is assumed that only three of the four binary matrices
($L^{[1]}$, $L^{[0]}$, $R^{[1]}$, and $R^{[0]}$) fit in the matrix buffers at
the same time to demonstrate the off-chip tiling capabilities. %
The P2S is not part of the example as it is assumed that the input matrix has
already been converted to the bit-serial layout. %
% Instructions
The corresponding instructions for each pipeline stage can be seen in
\autoref{tab:instruction_queues}, with $P$ denoting the matrix that accumulates
the result of these operations. %

% Fetch L1 and R1
The fetch stage begins by fetching $L^{[1]}$ and $R^{[1]}$ (instruction F1 and
F2) and then signals the execute stage (F3) that it can perform the first
binary-matrix multiplication (E2). %
% Compute L1xR1 while fetching L0
While the execute stage computes the dot product between $L^{[1]}$ and
$R^{[1]}$, the fetch stage continues fetching $L^{[0]}$, effectively achieving
an overlap between data fetch and execution (F4 and E2 performed in parallel). %
% Synchronization
Once the execute stage finishes the first binary-matrix multiplication, it
receives the signal from the fetch stage (F5) that $L^{[0]}$ resides in the
matrix buffers (E3). %
% Compute L0xR1
The execute stage continues by executing $L^{[0]} \cdot R^{[1]}$ (E4) while the
fetch stage has to wait since all the buffer space is occupied (F6). %
Note that E4 is part of a new wavefront, and the previous accumulator is
left-shifted by one by setting the appropriate accumulator mode to account for
this.
When the execute stage finishes the matrix multiplication, it signals the fetch
stage (E5). %
% Fetch R0 and then compute L1xR0 and L0xR0
Since $R^{[1]}$ is no longer needed, the fetch stage fetches $R^{[0]}$ (F7)
enabling the execute stage to finish the remaining matrix multiplications (E7
and E8). %
As this is the next step in the wavefront, this requires the accumulator to be
shifted again.
% Results
Once the execute stage has finished all binary matrix multiplications, it
signals the results stage (E9) which writes the result $P$ to main memory (R2). %

The schedule in \autoref{fig:timeline} causes the fetch stage and execute stage
to stall (F6 and E6) since there is not enough space to fetch $R^{[0]}$ before
$L^{[0]} \cdot R^{[1]}$ has been computed. %
An alternative schedule could be to split the binary matrices into tiles
enabling greater flexibility in what data to bring into the matrix buffers and
the possibility of overlapping fetch and execute. %

\begin{table}
  \caption{Initialized Instruction Queues for the Example Shown in \autoref{fig:bit-serial_example}}
  \label{tab:instruction_queues}
  \centering
  \begin{tabular}{l@{\hspace{0.3cm}}l@{\hspace{0.3cm}}l}
    \toprule
    Fetch & Execute & Result\\
    \midrule
% BEGIN RECEIVE ORGTBL alpha
F1 \RunInst{} L\(^{\text{[1]}}\) & E1 \GetToken{}  Fetch & R1 \GetToken{}  Execute\\
F2 \RunInst{} R\(^{\text{[1]}}\) & E2  \RunInst{} P = P +  L\(^{\text{[1]}}\)
                                  \mult{}  R\(^{\text{[1]}}\) & R2
                                                                \RunInst{} P \\
                                                              %  L \mult{}  R \\
F3  \PutToken{}  Execute & E3  \GetToken{}  Fetch & \\
F4  \RunInst{} L\(^{\text{[0]}}\) & E4  \RunInst{} P = (P<<1) + L\(^{\text{[0]}}\)\mult{}  R\(^{\text{[1]}}\) & \\
F5  \PutToken{}  Execute & E5  \PutToken{}  Fetch & \\
F6    \GetToken{}  Execute & E6  \GetToken{}  Fetch & \\
F7 \RunInst{} R\(^{\text{[0]}}\) & E7  \RunInst{} P = P + L\(^{\text{[1]}}\)\mult{} R\(^{\text{[0]}}\) & \\
F8 \PutToken{}  Execute & E8  \RunInst{} P = (P<<1) + L\(^{\text{[0]}}\)\mult{}  R\(^{\text{[0]}}\) & \\
 & E9 \PutToken{}  Result & \\
%END RECEIVE ORGTBL alpha
\bottomrule
  \end{tabular}
\end{table}

\section{Evaluation}

We implement the improved \OurScheme{} parametrizable hardware generator in
Chisel~\cite{bachrach2012chisel} and VHDL, and use Xilinx Vivado
2017.4~\cite{vivado_v2017.4} for synthesis, placement, and routing. %
We add registers to critical paths on the pipeline and enable register retiming
instead of manual floorplanning and timing optimizations to achieve higher clock
frequencies. %
We target the Ultra96 board~\cite{ultra96}, which has a Xilinx ZU3EG
MPSoC~\cite{zu3eg} containing an FPGA with 71k~LUTs and 214~BRAMs, and a
quad-core ARM Cortex-A53 CPU. %
The accelerator is connected to a 64-bit wide AXI high-performance port,
provisioning it with 4.8~GB/s of DRAM bandwidth when running at 300~MHz. %
The BISMO software stack and runtime are coded in C++, and executes on a single
ARM core. %
We use Ubuntu 18.04 provided with the PYNQ platform~\cite{pynq_doc:2018} for
Ultra96 as the operating system, and the PYNQ PMBUS interface for power
measurements. %
%It generates the instructions corresponding to chosen matrix dimensions, fills
%the instruction queues, places the bit-serial input matrices into DRAM and
%monitors the execution status of the accelerator. %
%We note that a soft processor such as MicroBlaze could also be used for this
%purpose, or the instructions can be simply pre-generated and placed into memory
%when the matrix dimensions to be multiplied are known ahead of time. %

As binary operations are the building block for bit-serial computations, we use
them as the common denominator for performance measurements. %
We treat \textsc{And} and popcount as analogues to multiplication and addition
when counting binary operations, i.e., a binary dot product between two
$N$-element binary vectors is counted as $2N$~binary operations. %

% \begin{table}
% 	\centering
% 	\caption{FPGA boards used for evaluation.}
% 	\begin{tabular}{lcccc}
% 	\toprule
% 	Board			& FPGA 				& LUT					& BRAM		&	DRAM bandwidth		\\
% 	\midrule
% 	PYNQ-Z1		& Z7020				& 53200				& 140			&	3.2~GB/s		\\
% 	ZC706			& Z7045				& 218600			& 545			&	12.8~GB/s		\\
% 	\bottomrule
% 	\end{tabular}
% 	\label{tab:boards}
% \end{table}

\subsection{Synthesis Results and Resource Cost}
\label{sec:synthesis}

We start by presenting synthesis results across a range of parameters for
different components of the \OurScheme{} architecture. %
Our aim is to explore the resource cost of scaling performance along different
axes of parallelism and building up a hardware cost model in the process. %
Unless otherwise stated, all data in this section is obtained by using
out-of-context synthesis for the ZU3EG FPGA, with a target clock period of 2~ns
to prioritize timing optimizations. %

\subsubsection{Dot Product Unit}
\label{sec:synthesis_DPU}

\begin{figure} [ht]
	\centering
	\resizebox{\linewidth}{!}{
		\begin{tikzpicture}
			\pgfplotstableread[col sep=tab]{data/dpu.txt}\dpudata;
			\begin{axis}[ylabel=Cost (LUT/bin.op.), axis y line*=right, ymin=0,
				width=\linewidth, height=4cm]
				\addplot [color=brown, dashed, mark=+] table [x=InpWidth, y=LUTPerOpOldDPU] {\dpudata};
				\addplot [color=blue, dashed, mark=*] table [x=InpWidth, y=LUTPerOpNewDPU] {\dpudata};
			\end{axis}
			\begin{axis}[xlabel=DPU width $D_k$ (bits),ylabel=Usage (LUT),
				%legend pos=north west,
                                legend style={at={(0.5, 1.1)},anchor=south},
				legend columns=2,
				width=\linewidth, height=4cm]
				\addlegendimage{brown, mark=+, only marks}
				\addlegendentry{$\mathrm{OpCost}_\mathrm{OldDPU}$}
				\addlegendimage{blue, mark=*, only marks}
				\addlegendentry{$\mathrm{OpCost}_\mathrm{NewDPU}$}
				\addplot [color=black,only marks] table [x=InpWidth, y=LUTOldDPUOldComp] {\dpudata};
				\addlegendentry{OldDPU}
				\addplot [black] table[y={create col/linear regression={y=LUTOldDPUOldComp}}] {\dpudata};
				\addlegendentry{%
				$\mathrm{LUT}_\mathrm{OldDPU}=2.04 \cdot D_k + 109 $ } %
				\addplot [color=red,only marks] table [x=InpWidth, y=LUTNewDPU] {\dpudata};
				\addlegendentry{NewDPU}
				\addplot [red] table[y={create col/linear regression={y=LUTNewDPU}}] {\dpudata};
				\addlegendentry{%
				$\mathrm{LUT}_\mathrm{NewDPU}=1.17 \cdot D_k + 44.1$ } %

			\end{axis}

		\end{tikzpicture}
	}
	\caption{DPU LUT usage and efficiency characterization.}
	\label{fig:dpuLUT}
\end{figure}
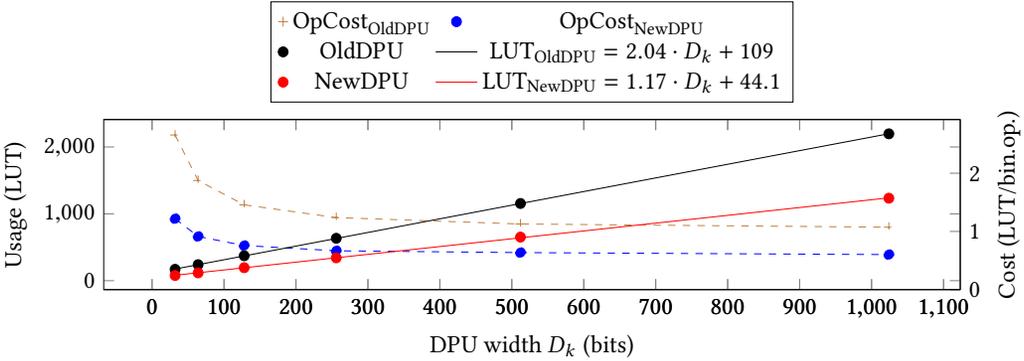

% Since the improved DPU does not include a standalone popcount module, we skip its
% characterization and provide results directly for the DPU.
We start by characterizing the resource cost of the DPU, which constitutes the
core computational unit of our overlay. %
\autoref{fig:dpuLUT} plots the LUT usage as well as the LUT cost per binary
operation of both the original and the improved DPUs. %
Similar to the original BISMO, the improved DPU resource cost includes the components
whose sizes is constant and does not scale with $D_k$, such as the accumulator
and mode multiplexer. %
We expect that their resource cost gets amortized for larger values of $D_k$,
making up a smaller proportion of the total DPU. %
The dashed lines in \autoref{fig:dpuLUT} plot the LUT cost per binary
operation. %
We observe that the cost per binary operation for the improved DPU starts at 1.2~LUTs
for $D_k=32$, decreasing to 0.6~LUTs for $D_k=1024$. %
Compared to the original BISMO with 2.6~LUTs for $D_k=32$ and 1.07~LUTs for
$D_k=1024$, this constitutes an improvement of $1.8\times$. %
Using linear regression on this data, the parameters $\mconst{DPU}$ and
$\aconst{DPU}$ of the \OurScheme{} cost model (\autoref{sec:costmodel_LUT}) are
1.17 and 44.1, respectively. %
We note that the additive constant $\aconst{DPU}$ for the improved DPU is 44.1
compared to 109 for the original DPU, decreasing the per-DPU overhead by 60\% due to
the removal of the expensive barrel shifter.  For the improved DPU, the reported
maximum frequency ($F_{\mathrm{max}}$) is between 600 and 719~MHz. %

\subsubsection{Fetch and Result Stage}
We evaluate the cost of the fetch and result stages for a single 64-bit memory
channel on the PYNQ-Z1, with $F$=$R$=64, $A$=32, and $B_r$=2. %
The fetch stage includes a DMA engine and the interconnect to move data into
matrix buffers. %
We observe that the LUT cost of the fetch stage is approximated well by
$1.89 \cdot (D_m + D_n) + 463$. %
We do not include the $1.89 \cdot (D_m + D_n)$ component in the cost model since
it is small even for large DPAs. %
The result stage includes a DMA engine, result matrix buffers, and a downsizer
(parallel-to-serial unit), which are all implemented using LUTs. %
The result buffer requires approximately $87.3 \cdot D_m \cdot D_n$~LUTs, while
the DMA engine and the downsizer need $32.8 \cdot D_m \cdot D_n + 255$~LUTs. %
Completing the cost model, the fetch and result stages contribute
$463 + 255 = 718$~LUTs to $\lutcost{base}$, which may increase with more
advanced DMA engines, and the LUT cost per DPU associated with the result stage
is $\lutcost{res} = 87.3 + 32.8 = 120.1$. %
%The DMA engine currently limits $F_{\mathrm{max}}$ to 200~MHz, and may be
%pipelined to further increase $F_{\mathrm{max}}$ for the entire accelerator.

\subsubsection{Parallel-to-Serial Accelerator Resource Cost}
\label{sec:p2scost}
To evaluate the cost of the hardware-accelerated data-layout conversion, we
evaluate the P2S with $M=8$ since 8-bit is the smallest natively supported
bit-parallel datatype for most CPUs. %
For $F=R=64$ the P2S contributes 929~LUTs to $\lutcost{base}$. %
Currently the majority of these LUTs are used for multiplexing between the
coalescing buffers when writing their contents to DRAM. %
As the access pattern to the coalescing buffers is quite regular, a more
optimized interconnect can be deployed here to further reduce the LUT cost. %

\subsubsection{Cost model validation}
\label{sec:synthesis_cost_model}

We generated 295 different \OurScheme{} designs ranging from ($D_m$=2, $D_k$=64,
$D_n$=2) to ($D_m$=12, $D_k$=256, $D_n$=10) in size to validate the cost models
described in \autoref{sec:cost_model}. %
The BRAM predictions were 100\% accurate for this particular range of designs. %
\autoref{fig:model_accuracy} shows the LUT usage from synthesis results versus
the prediction from the cost model. %
%We plot the LUT usage from synthesis results versus the prediction from the cost
%model in \autoref{fig:modelaccuracy}. %
The model's prediction is 97.8\% accurate on average across the tested sizes. %
\autoref{fig:model_accuracy_vs_size} shows how the prediction error is affected by
the size of the design. %
We observe that large designs are accurately predicted, while smaller designs
tend to be underestimated by the model.

\pgfplotstableread[col sep=tab]{data/accel_u96.txt}\cmvalidate

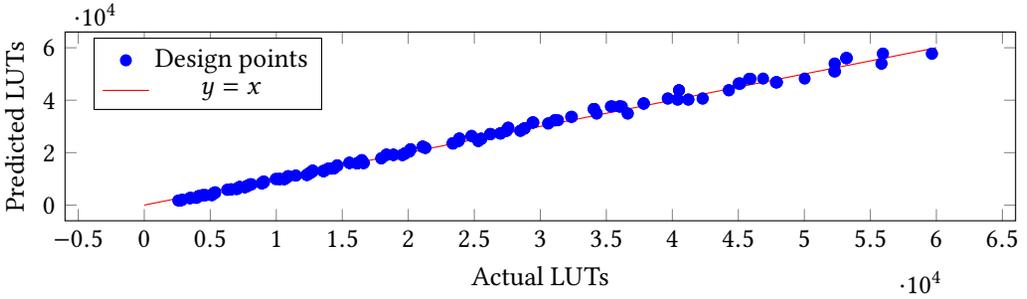
\begin{figure}
	\centering
	\resizebox{\linewidth}{!}{
		\begin{tikzpicture}
			\begin{axis}[xlabel=Actual LUTs ,ylabel=Predicted LUTs,
				legend pos=north west,
				width=\linewidth, height=4cm]
				\addplot [color=blue, only marks, mark=*] table [x=LUT, y=modelLUT] {\cmvalidate};
				\addlegendentry{Design points}
				\addplot [domain=0:60000, color=red]{x};
				\addlegendentry{$y=x$}
			\end{axis}
		\end{tikzpicture}
	}
	\caption{Predicted vs actual LUT usage.}
	\label{fig:model_accuracy}
\end{figure}

\begin{figure}
	\centering
	\resizebox{\linewidth}{!}{
		\begin{tikzpicture}
			\begin{axis}[xlabel=Actual LUTs,ylabel=Prediction Error \%,
				width=\linewidth, height=4cm, xmin=0,
                                xmax=65000, ymin=-42, ymax=22]
				\addplot [color=blue, only marks, mark=*] table [x=LUT, y=errpct] {\cmvalidate};
                                \addplot[domain=0:65000, color=red]{0};
				% \addlegendentry{}
			\end{axis}
		\end{tikzpicture}
	}
	\caption{LUT cost model prediction error with design size.}
	\label{fig:model_accuracy_vs_size}
\end{figure}
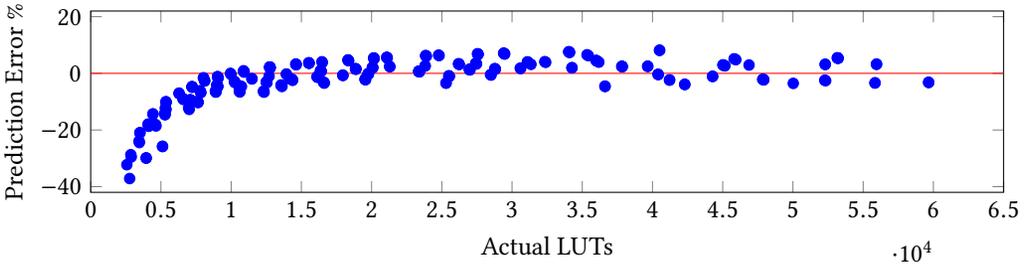

\subsubsection{LUT-BRAM Tradeoffs}
\autoref{fig:lutbramtradeoffs} shows three \OurScheme{} instances with the same
performance and buffer depth but different overlay dimensions
($D_m$, $D_k$, $D_n$) and plot the number of BRAMs used and the LUT cost per
binary operation. %
We observe a tradeoff between BRAM and LUT cost by scaling different
parameters. %
We see that larger $D_k$ results in lower LUT cost, but requires more BRAMs to
deliver the bandwidth. %
Conversely, smaller $D_k$ needs fewer BRAMs, but has larger LUT cost. %
We note that the DPA dimensions should be matched to the workload dimensions for
higher efficiency, e.g., $D_n > 1$ is wasteful for matrix-vector multiplication,
but LUT and BRAM budget may impose additional constraints. %

\pgfplotstableread[col sep=tab]{data/accel_tradeoffs_constperf.txt}\acceldata
\begin{figure}
	\centering
	\resizebox{\linewidth}{!}{
		\begin{tikzpicture}
			\begin{axis}[xtick=\empty, ylabel=LUT/bin.op., axis y line*=right, ymin=0,
				width=\linewidth, height=4cm]
				\addplot [color=red, mark=+] table [x=cname, y=LUT/op] {\acceldata};
			\end{axis}
			\begin{axis}[xlabel=Configuration {($D_m$, $D_k$, $D_n$)}, ylabel=BRAM,
                                legend style={at={(0.5, 0.72)},anchor=south},
				legend columns=3,
				xtick={1, 2, 3},
                                ymin=0,
				xticklabels={{(2, 1024, 2)}, {(4, 256, 4)}, {(8, 64, 8)}},
				width=\linewidth, height=4cm]
				\addplot [color=blue, mark=*] table [x=cname, y=BRAM] {\acceldata};
				\addlegendentry{BRAM}
				\addlegendimage{red}
				\addlegendentry{LUT/bin.op.}
			\end{axis}
		\end{tikzpicture}
	}
	\caption{LUT vs BRAM tradeoffs for 2.4~binary~TOPS at 300~MHz.}
	\label{fig:lutbramtradeoffs}
\end{figure}
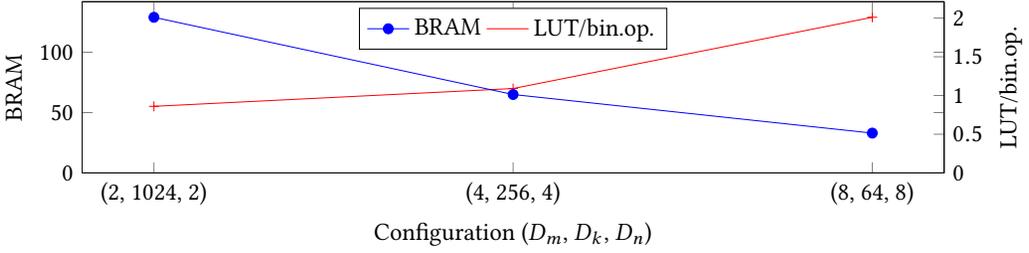

\begin{figure}
	\centering
	\resizebox{\linewidth}{!}{
		\begin{tikzpicture}
			\pgfplotstableread[col sep=tab]{data/multibitdpu.txt}\multibitdpudata;
			\begin{semilogxaxis}[xlabel=$w \times a$ dot products per cycle,ylabel=LUT/bin.op.,
				legend pos=north east, ymin=0.3, log basis x={2},
				legend style={font=\tiny},
				legend columns=5,
				width=\linewidth, height=4cm, cycle list name=Dark2-8]
				\addplot [index of colormap=0 of Dark2-8, mark=*, smooth] table [x=SIMD, y=bismo] {\multibitdpudata};
				\addlegendentry{\OurScheme{}}
				\addplot [index of colormap=1 of Dark2-8, mark=*, smooth] table [x=SIMD, y=a2b1] {\multibitdpudata};
				\addlegendentry{$2 \times 1$}
				\addplot [index of colormap=2 of Dark2-8, mark=*, smooth] table [x=SIMD, y=a2b2] {\multibitdpudata};
				\addlegendentry{$2 \times 2$}
				\addplot [index of colormap=3 of Dark2-8, mark=*, smooth] table [x=SIMD, y=a3b2] {\multibitdpudata};
				\addlegendentry{$3 \times 2$}
				%\addplot table [x=SIMD, y=a4b2] {\multibitdpudata};
				%\addlegendentry{4x2}
				\addplot [index of colormap=4 of Dark2-8, mark=*, smooth] table [x=SIMD, y=a3b3] {\multibitdpudata};
				\addlegendentry{$3 \times 3$}
			\end{semilogxaxis}
		\end{tikzpicture}
	}
	\caption{Comparing the LUT/bin.op. cost of bit-serial and bit-parallel DPUs.}
	\label{fig:dpumultibitLUT}
\end{figure}
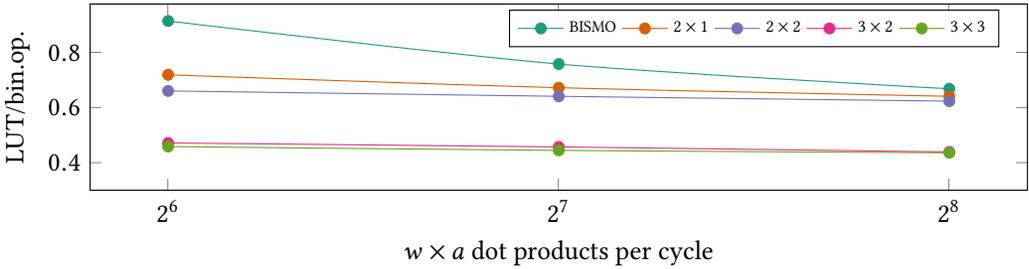

\subsubsection{Hardware Cost of Flexible Precision}

When the required precision is known beforehand, a matrix multiplier that uses
fixed-precision bit-parallel arithmetic is the commonly used alternative, though
bit-serial could still be used. %
To quantify the overhead associated with bit-serial for those cases, we
implemented a version of the DPU with $w \times a$-bit multipliers instead of
\textsc{And}, an adder tree instead of popcount, and no shifter and negator. %
This bit-parallel DPU performs the equivalent of $2 \cdot w \cdot a \cdot D_k$
binary operations per cycle using the same compressor generator as the BISMO
DPU, as explained in \autoref{sec:compressor}. %
\autoref{fig:dpumultibitLUT} compares the LUT cost for binary operation
equivalents between the \OurScheme{} DPU and several bit-parallel variants. %
We first observe that given the same %$D_k$,
number of bit-parallel operations ($w \cdot a$), the LUTs per binary operation
decreases with higher bit-parallel precision from 0.72 for $2 \times 1$ down to
0.46 for $3 \times 3$ when performing $2^6$ bit-parallel operations. %
As expected, bit-parallel DPUs have lower cost per bit operation compared to
bit-serial as they do not suffer from the shifter/negator overhead. %
For larger dot product sizes, the overhead is amortized and the worst-case gap
between \OurScheme{} and $3 \times 3$ closes down to 0.23 LUT per binary
operation. %
We note that this is not a fully fair comparison since \OurScheme{} hardware
supports significantly larger precisions compared to the fixed-precision
operators here. %
We also expect this data to be useful for designers who would like to build
\textit{digit-serial} architectures, where the building block can be, e.g.,
$2 \times 2$-bit matrix multipliers instead of binary. %

%\subsubsection{Scaling to Larger FPGAs}

%To better understand how \OurScheme{} would scale to larger FPGAs, we explore the LUT-limited scaling potential of the core compute element, the Dot Product Array (DPA).

\subsubsection{Scaling the \OurScheme{}  DPA to Larger Sizes}
\label{sec:scaling_DPA}

\looseness -1
\OurScheme{} scales performance by using a broadcast-style array of DPUs. %
Traditionally, semibroadcast or systolic arrays are preferred over broadcast for
VLSI designs due to the high fan-out requirements of broadcast
interconnects~\cite{kung1982systolic, zargham1996computer}. %
However, modern FPGAs have massive on-chip routing bandwidth and a large number
of flip flops for register duplication that can alleviate these concerns. %
To investigate how the \OurScheme{} DPA scales to larger sizes on Xilinx FPGAs,
we ran a number of experiments targeting the Xilinx Virtex UltraScale+
VU9P~\cite{vu9p} with out-of-context synthesis for large \OurScheme{} DPAs. %
Note that this assumes a LUT-bound design, i.e., we do not consider the matrix
buffer BRAMs necessary to feed the array, only the DPA itself. %
The results are summarized in \autoref{tab:large_dpa}. %
We observe that these large designs can still manage a respectable 500~MHz clock
without any manual floorplanning. %
The largest synthesized design uses approximately 80\% of the LUTs on this
device, achieving 783~binary TOPS at maximum frequency. %

\begin{table}
  \centering
  \caption{Large DPA synthesis results targeting Xilinx Virtex UltraScale+ VU9P.
  }
  \label{tab:large_dpa}
  \begin{tabular}{ccccccc}
    \toprule
		$D_m$ & $D_k$ & $D_n$ & LUT & FF & $F_{\mathrm{max}}$ (MHz) & Bin. TOPS \\
		\midrule
		50 & 64 & 50 & 337,500 & 555,000 & 523.56 & 167.54 \\
		16 & 1024 & 16 & 313,856 & 720,640 & 543.77 & 285.09 \\
		32 & 1024 & 16 & 627,712 & 1,441,280 & 532.77 & 558.64 \\
		32 & 1024 & 24 & 941,568 & 2,161,920 & 498.01 & 783.30 \\
    \bottomrule
  \end{tabular}
\end{table}

%  LocalWords:  pgfplotstableread popcountdata resizebox linewidth tikzpicture
%  LocalWords:  ylabel mathrm ymin addplot blue,mark PopCWidth fmax xlabel cdot
%  LocalWords:  Popcount addlegendimage addlegendentry red,only subsubsection
%  LocalWords:  pgfmathprintnumber pgfplotstableregressiona autoref bitwidth
%  LocalWords:  pgfplotstableregressionb Preu preusser2017generic dpudata xtick
%  LocalWords:  luteff fig:dpuLUT textsc mconst aconst sec:costmodel_LUT errpct
%  LocalWords:  bitwidths downsizer accelpynqdata LUTs,ylabel acceldata cname
%  LocalWords:  xticklabel scriptsize xticklabels multibitdpudata semilogxaxis
%  LocalWords:  cycle,ylabel bismo fig:dpumultibitLUT lutcost textit
%  LocalWords:  fig:lutbramtradeoffs cmvalidate psosixtyfour Virtex sec:p2scost
%  LocalWords:  psoonehundred psottwohundred psofivehundred toprule
%  LocalWords:  fig:p2s_model_accuracy_vs_size zargham1996computer
%  LocalWords:  floorplanning tab:large_dpa midrule bottomrule semibroadcast

\subsection{Runtime Performance}
\label{sec:performance}

In this section, we assess the runtime performance and energy efficiency
achievable by the improved \OurScheme{} instances running on the Ultra96. %
We assume that the input matrices are stored in DRAM using a bit-packed data
layout (\autoref{sec:bsdatalayout}) %\cite{umuroglu_jahre:CASES2017},
and that one matrix is transposed. %
We create matrix-multiplication workloads with different dimensions and
bitwidths, manually build the corresponding instruction sequences, and run the
workloads on the enumerated \OurScheme{} instances listed in
\autoref{tab:runtime_hw_configs} to evaluate how the overlay size interacts with
workload size. %
We also reproduce the original \OurScheme{} results in
\autoref{tab:runtime_hw_configs_old} to demonstrate the improvements of the new
BISMO. %
For instance, the $8 \times 256 \times 8$ instance is 27\% smaller for the
improved \OurScheme{}, and the design can be clocked $1.5\times$ faster compared
to the original. %
The resource improvement is due to the improved DPU design as the LUTs
themselves are very similar between the two devices, while the clock improvement
mainly comes from the process node improvement (16 vs 28~nm). %

% To gain insight into how the overlay dimensions interact with workload size,
% we create matrix multiplication workloads with different dimensions and
% bitwidths, manually build the corresponding instruction sequences, and then
% run the workloads on the enumerated \OurScheme{} instances listed in
% \autoref{tab:runtime_hw_configs}. %

\begin{table}
  \centering
  \caption{Improved \OurScheme{} instances for runtime measurements on the Ultra96.
  }
  \label{tab:runtime_hw_configs}
  \begin{tabular}{cccccccc}
    \toprule
    \# & $D_m$ & $D_k$ & $D_n$ & LUT & BRAM & $F_{\mathrm{max}}$ (MHz) & GOPS \\
    \midrule
    1 & 4 & 256 & 4 & 12,657 (18\%) & 65 (31\%) & 313.19 & 2,565.6 \\
    2 & 8 & 256 & 4 & 19,613 (28\%) & 97 (46\%) & 323.31 & 5,297.1 \\
    3 & 8 & 256 & 8 & 33,418 (48\%) & 129 (61\%) & 309.89 & 10,154.3 \\
    4 & 10 & 128 & 10 & 34,252 (49\%) & 81 (38\%) & 306.84 & 7,855.2 \\
    5 & 12 & 256 & 6 & 36,879 (53\%) & 145 (68\%) & 302.39 & 11,147.3 \\
    6 & 12 & 128 & 12 & 46,847 (67\%) & 97 (46\%) & 281.85 & 10,390.1 \\
    7 & 10 & 256 & 10 & 50,734 (72\%) & 161 (76\%) & 311.53 & 15,950.2 \\
    %8 & 12 & 256 & 12 & 68145 (97\%) & 193 (91\%) & 149.72 & 11,038.78 \\
    \midrule
    \multicolumn{8}{c}{$F=R=64$ and $F_{\mathrm{clk}}=300~\mathrm{MHz}$.} \\
    \bottomrule
  \end{tabular}
\end{table}

\begin{table}
  \centering
  \caption{Original \OurScheme{} instances for runtime measurements on the
  PYNQ-Z1~\cite{umuroglu+:FPL2018}.
  }
  \label{tab:runtime_hw_configs_old}
  \begin{tabular}{cccccc}
    \toprule
    $D_m$ & $D_k$ & $D_n$ & LUT & BRAM & GOPS \\
    \midrule
    8 & 64 & 8 & 19,545 (37\%) & 121 (86\%) & 1,638.4 \\
    8 & 128 & 8 & 27,740 (52\%) & 129 (92\%) & 3,276.8 \\
    8 & 256 & 8 & 45,573 (86\%) & 129 (92\%) & 6,553.6 \\
    4 & 256 & 4 & 13,352 (25\%) & 129 (92\%) & 1,638.4 \\
    8 & 256 & 4 & 24,202 (45\%) & 129 (92\%) & 3,276.8 \\
    4 & 512 & 4 & 21,755 (41\%) & 129 (92\%) & 3,276.8 \\
    \midrule
    \multicolumn{6}{c}{$F=R=64$ and $F_{\mathrm{clk}}=200~\mathrm{MHz}$.} \\
    \bottomrule
  \end{tabular}
\end{table}

\subsubsection{Peak Binary Compute}
\label{sec:peakcompute_bin}
%As binary operations are the building blocks for bit serial,
We start by measuring the maximum achievable binary matrix-multiply performance
dictated purely by the execute stage. %
For this experiment, we assume that the matrices have already been fetched into
on-chip memory and disregard the cost of result writing. %
\autoref{fig:execeff} plots the achieved performance for different number of
matrix columns ($K$) as a percentage of observed peak performance for different
popcount widths ($D_k$). %
% \autoref{fig:execeff} plots the percentage of peak performance achieved by
% on-chip workloads with different number of columns. %
We observe that the efficiency increases with more columns, and that instances
with larger $D_k$ require wider matrices than smaller $D_k$ ones to be
efficient. %
As an example, for a matrix with 8192 columns (dotted line in
\autoref{fig:execeff}), the instance with $D_k=256$ reaches 68\%
efficiency, while $D_k=128$ achieves 82\%. %
Wide matrices achieve close to 100\% of the peak performance for all instances.
The inefficiency for narrow matrices is due to the lack of work to fill the
compressor pipeline.
For example, assume that a $D_k=1024$ compressor pipeline has 10 stages, and is
processing a dot product with $K=6144$.
This workload is fed to the compressor within 6 clock cycles, after which the
execute stage controller must wait for the operation to complete to synchronize with
the result stage, thus creating bubbles in the pipeline.
This can be remedied by decreasing the DPA pipeline depth. %
As the improved \OurScheme{} DPU has fewer compressor stages compared to the
original \OurScheme{}~\cite{umuroglu+:FPL2018}, we observe up to 10\% relative
improvement for the same $D_k$ and matrix size.

\pgfplotstableread[col sep=tab]{data/execeff_64.txt}\execeffs
\pgfplotstableread[col sep=tab]{data/execeff_128.txt}\execeffm
\pgfplotstableread[col sep=tab]{data/execeff_256.txt}\execeffl
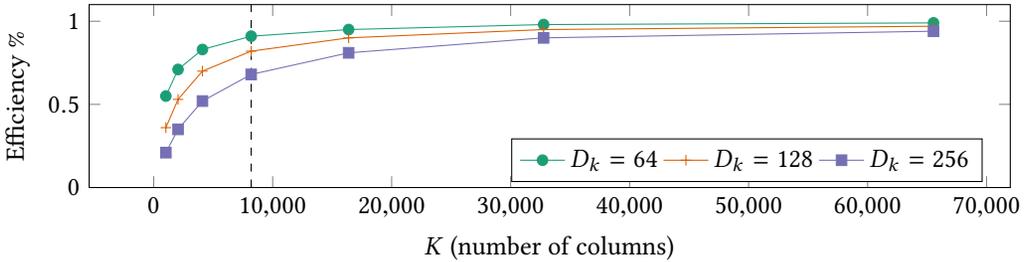
\begin{figure}
	\centering
	\resizebox{\linewidth}{!}{
		\begin{tikzpicture}
			\begin{axis}[xlabel=$K$ (number of columns),ylabel=Efficiency \%,
				legend columns=3, scaled x ticks=false,
                                legend pos= south east,
                                ymin=0, ymax=1.1,
				width=\linewidth, height=4cm]
				\addplot[mark=*, index of colormap=0 of Dark2-8] table [x=K, y=ExecEff] {\execeffs};
				\addlegendentry{$D_k=64$}
        \addplot[mark=+, index of colormap=1 of Dark2-8] table [x=K, y=ExecEff] {\execeffm};
				\addlegendentry{$D_k=128$}
        \addplot[mark=square*, mark options={solid}, index of colormap=2 of Dark2-8] table [x=K, y=ExecEff] {\execeffl};
				\addlegendentry{$D_k=256$}
         \draw [dashed] (8192,0) -- (8192,1.1);
                        \end{axis}
		\end{tikzpicture}
	}
	\caption{Execute stage efficiency depending on $D_k$ and number of columns $K$.}
	\label{fig:execeff}
\end{figure}

\subsubsection{Peak Bit-Serial Compute}
\label{sec:peakbitserial}
Per Algorithm~\ref{alg:bit-serial_matrix_multiplication}, if the runtime of a
binary ($1 \times 1$) matrix multiplication of a given size is $t$, we expect
the runtime of a $w \times a$-bit matrix multiplication of the same size to be
$w \cdot a \cdot t$. %
\autoref{fig:multibit} plots the performance for matrices of size
$10 \times 2048 \times 10$ and $10 \times 16384 \times 10$ with increasing
$w, a$ on instance~\#4. %
We observe slightly better performance than the projected $w \cdot a \cdot t$
since multiple dot products are accumulated together for the multi-bit case,
behaving like a longer dot product and increasing the execute-stage efficiency
(\autoref{fig:execeff}). %

\pgfplotstableread[col sep=tab]{data/multibit_k128_K2048.txt}\multibitressmall
\pgfplotstableread[col sep=tab]{data/multibit_k128_K16384.txt}\multibitreslarge
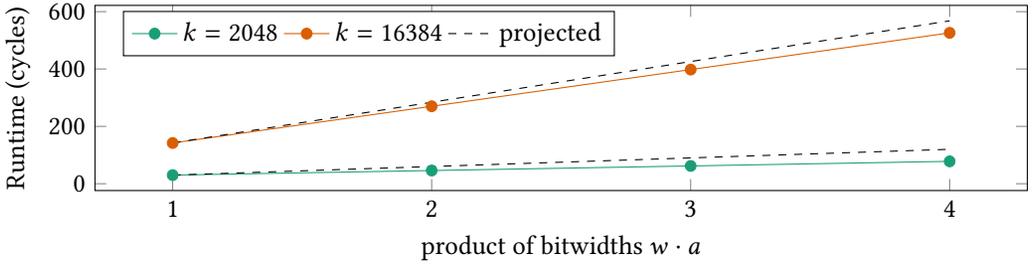
\begin{figure}
	\centering
	\resizebox{\linewidth}{!}{
		\begin{tikzpicture}
			\begin{axis}[xlabel=product of bitwidths $w \cdot a$, ylabel=Runtime (cycles),
        xtick=data,
                                legend pos=north west,
				legend columns=3,
				width=\linewidth, height=4cm]
				\addplot[mark=*, index of colormap=0 of Dark2-8] table [x=bitcomp, y=cycles] {\multibitressmall};
				\addlegendentry{$k=2048$}
        \addplot[mark=*, index of colormap=1 of Dark2-8] table [x=bitcomp, y=cycles] {\multibitreslarge};
				\addlegendentry{$k=16384$}
        \addplot[style=dashed] table [x=bitcomp, y=proj_bin] {\multibitreslarge};
        \addplot[style=dashed] table [x=bitcomp, y=proj_bin] {\multibitressmall};
        \addlegendentry{projected}
			\end{axis}
		\end{tikzpicture}
	}
	\caption{Runtime with increasing precision on instance \#4.}
	\label{fig:multibit}
\end{figure}

\subsubsection{Stage Overlap}
We now quantify the performance gain by overlapping the fetch, execute, and
result stages for larger matrix multiplications. %
Using the block matrix multiplication algorithm from Matam
et~al.~\cite{matam2013energy} we create an instruction sequence to run a
$256 \times 4096 \times 256$ binary matrix multiplication on an
$8 \times 64 \times 8$ instance.
The input matrices here are twice the size of the on-chip memory,
similar to the example in \autoref{sec:schedexample}. %
By overlapping the operation of different stages, the multiplication finishes in
121,133 cycles, achieving a speedup of $2.2\times$ compared to the 266,510
cycles when the stages are executing without overlap. %

\subsubsection{Power Consumption}

%
% POWER TABLE
%
\begin{table}
\centering
  \caption{Power consumption data from improved \OurScheme{} on the Ultra96.}
  \label{tab:power_results}
  \begin{tabular}{crrrrcc}
    \toprule
    Configuration & \multicolumn{4}{c}{Power (W)} & Binary & Binary \\
    \cline{2-5}
    $(\mathrm{Instance}, F_{\mathrm{clk}})$ & Idle & Exec & F \& R & Full & GOPS & GOPS/W \\
    \midrule
    %(8x256x8, 50 MHz) & 4.98 & +0.01 & +0.25 & 5.26 & 1,638.40 & 311.72 \\
    %12x128x12, 50 MHz & 4.99 & +0.12 & +0.17 & 5.32 & 1,843.20 & 346.73 \\
    (10x256x10, 50 MHz) & 5.10 & +0.01 & +0.26 & 5.39 & 2,560.00 & 475.13 \\
    (4x256x4, 300 MHz) & 5.39 & +0.09 & +0.30 & 5.76 & 2,457.60 & 426.67 \\
    %4x256x4, 50 MHz & 4.76 & +0.02 & +0.29 & 5.09 & 409.60 & 80.50 \\
    %8x256x4, 50 MHz & 4.88 & +0.01 & +0.23 & 5.15 & 819.20 & 159.13 \\
    %10x128x10, 50 MHz & 4.92 & +0.05 & +0.24 & 5.20 & 1,280.00 & 246.34 \\
    %8x256x4, 300 MHz & 5.74 & +0.11 & +0.36 & 6.11 & 4,915.20 & 804.72 \\
    %10x128x10, 300 MHz & 5.83 & +0.11 & +0.32 & 6.22 & 7,680.00 & 1,235.52 \\
    %12x128x12, 300 MHz & 6.20 & 0.17 & 0.38 & 6.64 & 11,059.20 & 1,666.55 \\
    \midrule
    (8x256x8, 300 MHz) & 6.17 & +0.17 & +0.41 & 6.65 & 9,830.40 & 1,478.70 \\
    (10x256x10, 300 MHz) & 6.76 & +0.23 & +0.36 & 7.20 & 15,360.00 & 2,133.33 \\
    \bottomrule
  \end{tabular}
\end{table}
\begin{table}
\centering
  \caption{Power consumption data from the original \OurScheme{} instances on
  PYNQ-Z1~\cite{umuroglu+:FPL2018}.}
  \label{tab:power_results_old}
  \begin{tabular}{crrrrcc}
    \toprule
    Configuration & \multicolumn{4}{c}{Power (W)} & Binary & Binary \\
    \cline{2-5}
    $(\mathrm{Instance}, F_{\mathrm{clk}})$ & Idle & Exec & F \& R & Full & GOPS & GOPS/W \\
    \midrule
    (8x64x8, 200 MHz) & 2.53 & +0.33 & +1.09 & 4.07 & 1,638.00 & 402.16 \\
    (8x128x8, 100 MHz) & 2.10 & +0.19 & +0.87 & 3.11 & 1,638.00 & 527.51 \\
    (8x256x8, ~50 MHz) & 1.76 & +0.30 & +0.63 & 2.53 & 1,638.00 & 646.39 \\
    (4x256x4, 200 MHz) & 2.53 & +0.34 & +1.09 & 3.86 & 1,638.00 & 424.98 \\
    (8x256x4, 100 MHz) & 2.05 & +0.24 & +0.92 & 3.06 & 1,638.00 & 536.02 \\
    \midrule
    (4x512x4, 200 MHz) & 2.87 & +0.71 & +1.19 & 4.64 & 6,554.00 & 1,413.39 \\
    \bottomrule
  \end{tabular}
\end{table}

We use the PMBUS interface on the Ultra96 to measure the total board power while
running one or more stages in a loop to measure the power efficiency of
\OurScheme{}. %
We turn off the wireless interfaces on the Ultra96 to obtain better idle power
readings. %
\autoref{tab:power_results} lists the power consumption of four instances. %
In the top part of the table, we compare three different design points with
similar performance, while the bottom part are top-performance designs. %
We list four power readings: %
% Idle
the idle power with no stages running,
% execute
the increment from idle with only the execute stage running,
% fetch and result
the increment with only the fetch and result stages running, and
% full
the full power with all stages running. %

Overall, the idle power on the Ultra96 constitutes more than 90\% of the full
power consumption. %
We find that on average the execute stage contributes 1\% of the full power
consumption, while the fetch and result stages contribute 5\%. %
For the cases with similar performance, we see that a large but slow-clocked
design achieves $1.1\times$ better power efficiency than a small but
fast-clocked design, similar to what is reported for
FINN~\cite{umuroglu+:FPGA2017finn}. %
%The majority of this increase in power efficiency can be attributed to lower
%idle power due to a slower clock.
% comparison to old BISMO
We also include the original \OurScheme{} power consumption
data %~\cite{umuroglu+:FPL2018}
in \autoref{tab:power_results_old} for comparison. %
Although the Ultra96 has higher idle power consumption compared to the PYNQ-Z1,
we see that the improved \OurScheme{} has 1.5$\times$ better power efficiency
compared to the original \OurScheme{} for the top-performing designs. %
This can be attributed to a combination of process scaling (16 vs 28~nm) and the
more LUT-efficient design in the improved \OurScheme{}. %

\subsubsection{Parallel-to-Serial Accelerator Performance}
\label{sec:p2s_performance}

To quantify the performance gains from the P2S accelerator, we compare the
execution time for data layout transformation between the accelerator (with the
parameters in \autoref{sec:p2scost}) and a CPU version on the Ultra96.
For the CPU version, we use the open-source implementation
from~\cite{umuroglu_jahre:CASES2017}.
This is a single-thread implementation that uses 32-bit multiplication with
a specifically crafted constant to pack bit positions from multiple 8-bit words,
originally proposed by Mula~\cite{mula+:bitpack}.
We report the average of 30 runs to account for caching effects.
As the Ultra96 ZU3EG possesses a 64-bit quad-core CPU, we optimistically divide
the CPU execution time by eight to allow for future multithreading and wider
datapath optimizations.

\begin{figure}
   \centering
   %\resizebox{\linewidth}{!}{
     \begin{tikzpicture}
     \begin{axis}[
         ybar,
         enlargelimits=0.15,
         legend pos=north west,
         ylabel={Execution Time (us)},
         symbolic x coords={1-bit, 2-bit, 3-bit, 4-bit},
         xtick=data,
         nodes near coords,
         nodes near coords align={vertical},
         width=\linewidth, height=6cm
         ]
     %\addplot coordinates {(tool8,7) (tool9,9)};
     %\addplot coordinates {(tool8,4) (tool9,4)};
     %\addplot coordinates {(tool8,1) (tool9,1)};
    \addplot coordinates {(1-bit,185.8) (2-bit,274.7) (3-bit,363.6) (4-bit,452.5)};
    \addplot coordinates {(1-bit,18.5) (2-bit,21.1) (3-bit,24.8) (4-bit,28.3)};
    %\addplot coordinates {(2-bit,2197.70) (2-bit,21.15)};
    %\addplot coordinates {(3,2908.58) (3,24.85)};
    %\addplot coordinates {(4,3619.81) (4,28.27)};
    %\addplot coordinates {(5,4396.29) (5,27.685)};
    %\addplot coordinates {(6,5048.97) (6,30.355)};
    %\addplot coordinates {(7,5756.30) (7,38.35)};

     \legend{Cortex-A53, \OurScheme{} P2S}
     \end{axis}
     \end{tikzpicture}
   %}
   \caption{Parallel-to-serial runtime for 20x1280 matrices of varying precision.}
   \label{fig:p2sbenchmark}
 \end{figure}
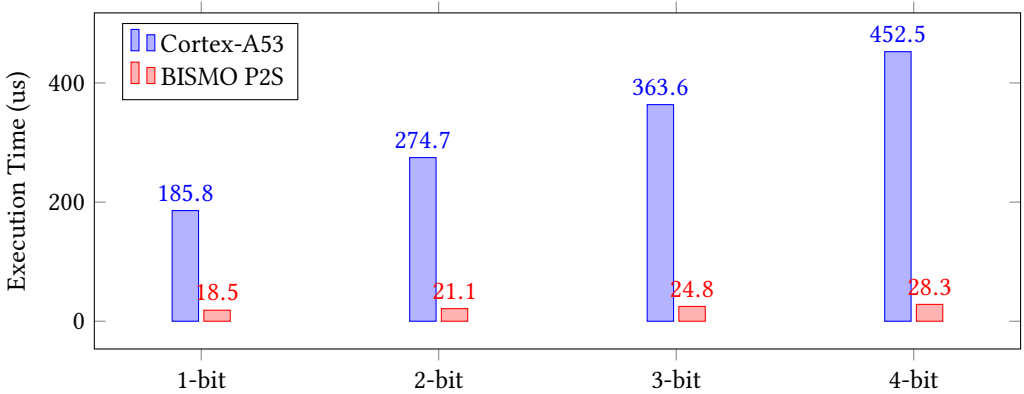

\autoref{fig:p2sbenchmark} plots the execution time for both methods for a
20x1280 matrix of varying precision stored using 8-bit elements.
On average, the P2S accelerator is $13.8\times$ faster than the CPU
implementation.
The CPU is limited by its ability to perform fine-grained (bit-level) data
movement between registers, while the FPGA is well-suited to this task.
Especially for larger matrices where data layout conversion can become costly,
the P2S accelerator can contribute significantly to overall bit-serial matrix
multiplication peformance.
%For platforms with distinct (not coherent) CPU and FPGA address spaces, the P2S
%accelerator has the added benefit of avoiding copy operations between different
%address spaces.

%\pgfplotstableread[col sep=comma]{data/p2sbenchmarking.txt}\benchmarkultraone
%\pgfplotstableread[col sep=tab]{data/p2s_bench_7_bit.txt}\benchmarkultraseven

%  LocalWords:  subsubsection sec:peakcompute_bin fig:execeff ylabel
%  LocalWords:  pgfplotstableread execeffs execeffm execeffl addplot
%  LocalWords:  resizebox linewidth tikzpicture ExecEff cdot xtick
%  LocalWords:  addlegendentry sec:peakbitserial multibitressmall
%  LocalWords:  alg:bit-serial_matrix_multiplication multibitreslarge
%  LocalWords:  bitwidths bitcomp proj_bin umuroglu scriptsize vspace
%  LocalWords:  crrrrcc umuroglu_jahre:CASES2017 tab:relatedwork emph
%  LocalWords:  ccccrrc multirow rotatebox Pedersoli judd2016stripes
%  LocalWords:  pedersoli2017espresso sec:schedexample
%  LocalWords:  benchmarkultraone benchmarkultraseven
%  LocalWords:  fig:p2sbenchmark

%  LocalWords:  bachrach2012chisel Vivado autoref Zynq textsc popcount toprule
%  LocalWords:  midrule bottomrule tikzpicture xlabel ylabel linewidth addplot
%  LocalWords:  pgfplotstableread popcountdata addlegendentry cdot mathrm ymin
%  LocalWords:  pgfmathprintnumber pgfplotstableregressiona subsubsection
%  LocalWords:  pgfplotstableregressionb resizebox PopCWidth luteff red,only
%  LocalWords:  addlegendimage fig:dpuLUT floorplanning zu3eg pynq_doc:2018

\section{Related Work}
\label{sec:relatedwork}

% SoTA TABLE
\begin{table*}
  \caption{Comparing \OurScheme{} to recent work.}
  \label{tab:relatedwork}
  \centering
  \resizebox{\columnwidth}{!}{
  \begin{tabular}{ccccrrc}
    \toprule
    Work & Platform & Type & Precision & Binary GOPS & GOPS/W & \\
    \midrule
    Improved \OurScheme{} & ZU3EG on Ultra96 & FPGA & bit-serial & 15,360 & 2,133.33 &
    \parbox[t]{2mm}{\multirow{6}{*}{\rotatebox[origin=c]{90}{incl. DRAM}}} \\
    Original \OurScheme{} \cite{umuroglu+:FPL2018} & Z7020 on PYNQ-Z1 & FPGA & bit-serial & 6,554 & 1,413.40 & \\
    FINN \cite{umuroglu+:FPGA2017finn} & Z7045 on ZC706 & FPGA & binary & 11,613 & 407.50 & \\
    Moss et al. \cite{moss2018customizable} & GX1150 on HARPv2 & FPGA & reconfigurable & 41 & 849.38 & \\
    Umuroglu et al. \cite{umuroglu_jahre:CASES2017}$\dagger$ & Cortex-A57 on Jetson TX1 & CPU & bit-serial & 92 & 18.80 & \\
    Pedersoli et al. \cite{pedersoli2017espresso}$\dagger$ & GTX 960 & GPU & limited bit-serial & 90,909 & 757.60 & \\
    Judd et al. \cite{judd2016stripes}$\dagger$ & ASIC & ASIC & limited bit-serial & 128,450 & 4,253.30 & \\
    \midrule
    Improved \OurScheme{} & ZU3EG on Ultra96 & FPGA & bit-serial & 15,360 & 2,245.61 &
    \parbox[t]{2mm}{\multirow{4}{*}{\rotatebox[origin=c]{90}{excl. DRAM}}} \\
    Original \OurScheme{} \cite{umuroglu+:FPL2018} & Z7020 on PYNQ-Z1 & FPGA & bit-serial & 6,554 & 1,889.70 & \\
    FINN \cite{umuroglu+:FPGA2017finn} & Z7045 on ZC706 & FPGA & binary & 11613 & 992.50 & \\
    Umuroglu et al. \cite{umuroglu_jahre:CASES2017}$\dagger$ & Cortex-A57 on Jetson TX1 & CPU & bit-serial & 92 & 43.80 & \\
    Umuroglu et al. \cite{umuroglu_jahre:CASES2017}$\dagger$ & i7-4790 & CPU & bit-serial & 355 & 12.20 & \\
    %\midrule
    \multicolumn{7}{c}{\textit{$\dagger$ indicates our experiments from released code or projections based on paper.}} \\
    \bottomrule
  \end{tabular}
  }
\end{table*}

\autoref{tab:relatedwork} compares \OurScheme{} against several
recently-proposed implementations for low-precision matrix
multiplication, % on different platforms,
using peak binary performance and performance per watt as metrics. %
The top part of the table includes DRAM power, while the bottom part only
considers on-chip compute and memory power. %
The improved \OurScheme{} presented in this work achieves a peak energy
efficiency of 2.13 binary TOPS/W, which is an improvement of $1.5\times$ compared to
the original \OurScheme{}~\cite{umuroglu+:FPL2018}. %
The peak performance of improved \OurScheme{} is also $2.3\times$ that of the
original, owing to a combination of the improved DPU design and newer FPGA.
To our knowledge, \OurScheme{} is the first FPGA implementation for bit-serial
matrix multiplication, but comparable related work on binarized neural networks
by Umuroglu et al.~\cite{umuroglu+:FPGA2017finn} and low-precision matrix
multiplication by Moss et al.~\cite{moss2018customizable} report respectively
$5.2\times$ and $2.5\times$ lower power efficiency than ours. %
Although the GPU binary matrix multiplication kernels proposed by Pedersoli et
al.~\cite{pedersoli2017espresso} achieve an impressive 90~TOPS for large binary
matrices, their work does not report power measurements. %
Assuming a power consumption of 120~W for the GTX 960, \OurScheme{} achieves
$2.8\times$ better power efficiency in comparison. %
On CPUs, the single-threaded implementation by Umuroglu and
Jahre~\cite{umuroglu_jahre:CASES2017} performed far worse than \OurScheme{}, and
is still outperformed by more than an order of magnitude even when assuming
$4\times$ performance improvement with multi-core parallelization. %
Finally, Stripes by Judd et al.~\cite{judd2016stripes} outperforms ours by
$2.0\times$ due to the performance and energy efficiency of an ASIC
implementation. %

%  LocalWords:  sec:relatedwork tab:relatedwork resizebox columnwidth ccccrrc
%  LocalWords:  toprule midrule multirow rotatebox umuroglu Pedersoli textit
%  LocalWords:  umuroglu_jahre:CASES2017 pedersoli2017espresso judd2016stripes
%  LocalWords:  bottomrule

\section{Conclusion}

We have presented an improved version of \OurScheme{}, a bit-serial matrix
multiplication overlay that can scale its precision to match an application's
computational requirements and its hardware to match available system
resources. %
A new architecture and an FPGA specific compressor implementation for the dot
product unit (DPU) are shown to reduce the LUT cost per binary operation
by $1.8\times$ compared to the original \OurScheme{}. %
The new design achieves a peak performance of 15.4~binary~TOPS with an energy
efficiency of 2.1~TOPS/W on an Ultra96 board, an improvement of $2.3\times$ and
$1.5\times$, respectively. %
Synthesis results targeting a Xilinx Virtex UltraScale+ VU9P show that the core
dot product array (DPA) can achieve a peak performance of 783~binary~TOPS at
500~MHz and a LUT utilization of 80\%. %

% Cost model
% The proposed cost model accurately predicts FPGA resource utilization and enables
% quick performance estimations. %
% Software
% \OurScheme{} is software programmable, providing the possibility to adapt its
% execution to the dimension and precision of any input matrix. %
% Performance
% Our evaluation indicates that the improved \OurScheme{} achieves a peak
% performance of 15.4 TOPS with an energy efficiency of up to 2.1 TOPS/W on an
% Ultra96 board. %

% Here we have demonstrated that \OurScheme{} generates an energy efficient
% overlay for bit-serial matrix multiplication. %
% The relationship between the bit-serial matrix multiply algorithm and the
% generated hardware is depicted to show the correctness. %
% The usage of software instructions to enable flexibility to the programmer
% is shown.

% \subsection{Future Work}

% In the future we would like to investigate the properties with regard to
% different schedules depending on the matrix. %
% Different matrix workloads may prefer different overlays generated
% by \OurScheme{}. %
% Exploring the optimum design space between the workload, generated hardware and
% the possible optimum schedule is one area we would like to look at.

% In addition to this, we would like explore the possibility of having arbitrary
% bit-precision matrix multiplication, which can enable approximate computing.

% HW/SW co-design/DSE; different matrix workloads may prefer different HW dimensions.
% explore applications for bit skipping/bit sparse.

%\section*{Acknowledgments}

%  LocalWords:  Virtex

\section*{Acknowledgments}

This work was funded by Vetenskapsr\r{a}det project 2015-05159. %
The computations were performed on resources provided by NTNU through the EPIC cluster. %

\bibliographystyle{ACM-Reference-Format}
\bibliography{defs,refs}

\end{document}